\documentclass[12pt, journal,onecolumn]{IEEEtran}

\usepackage{amssymb,amsmath,epsfig}
\usepackage{cite}
\usepackage{graphicx}
\usepackage{algorithmic}
\usepackage{algorithm}
\usepackage{amsfonts}
\usepackage{amstext}
\usepackage{graphicx}
\usepackage{setspace}
\usepackage{amssymb}
\usepackage{esint}
\usepackage{amsthm}
\usepackage{perpage} %the perpage package
\usepackage{epstopdf}
\usepackage{mathtools}
\usepackage{verbatim}
\usepackage{flushend}
\usepackage{color}
\usepackage[pagewise]{lineno}
\AtEndDocument{\par\leavevmode}
\renewcommand{\baselinestretch}{2}

\begin{document}

\title{Adaptive Polar Active Contour for Segmentation and Tracking in Ultrasound Videos}

\author{Ebrahim~Karami,~\IEEEmembership{Member,~IEEE}, Mohamed~S.~Shehata,~\IEEEmembership{Senior~Member,~IEEE}, and Andrew~Smith
\thanks{Manuscript received December 3, 2017; revised February 19, 2018; accepted March 5, 2018. The review of this paper was coordinated by Prof. Alexander C. Loui.}
\thanks{E. Karami and M. S. Shehata are with the Faculty of Engineering and Applied Sciences, Memorial University, St. John's, Canada email: \{ekarami, mshehata\}@mun.ca.}

\thanks{A. Smith is with Faculty of Medicine, Memorial University, St. John's, Canada, email: andrew.smith@med.mun.ca.}
\thanks{Copyright ©2018 IEEE. Personal use of this material is permitted. However, permission to use this material for any other purposes must be obtained from the IEEE by sending an email to pubs-permissions@ieee.org.}}

%\author{\IEEEauthorblockN{Ebrahim~Karami\IEEEauthorrefmark{1}, Mohamed~S.~Shehata\IEEEauthorrefmark{1}\IEEEauthorrefmark{3}, Peter McGuire\IEEEauthorrefmark{2}, and Andrew Smith\IEEEauthorrefmark{1}\IEEEauthorrefmark{3}}
%\\\IEEEauthorblockA{\IEEEauthorrefmark{1}Faculty of Engineering and Applied Sciences,\\ Memorial University, St. John's, Canada}
%\\\IEEEauthorblockA{Email: \{ekarami,mshehata\}@mun.ca}
%\\\IEEEauthorblockA{\IEEEauthorrefmark{2}C-CORE, St. John's, Canada}
%\\\IEEEauthorblockA{Email: peter.mcguire@c-core.ca}
%\\\IEEEauthorblockA{\IEEEauthorrefmark{3}Faculty of Medicine, Memorial University, St. John's, Canada}
%\\\IEEEauthorblockA{Corresponding Author Email: ekarami@mun.ca}}

\maketitle
%\IEEEpeerreviewmaketitle

\begin{abstract}
Detection of relative changes in circulating blood volume is important to guide resuscitation and manage a variety of medical conditions including sepsis, trauma, dialysis and congestive heart failure. Recent studies have shown that estimates of circulating blood volume can be obtained from the cross-sectional area (CSA) of the internal jugular vein (IJV) from ultrasound images. However, accurate segmentation and tracking of the IJV in ultrasound imaging is a challenging task and is significantly influenced by a number of parameters such as the image quality, shape, and temporal variation. In this paper, we propose a novel adaptive polar active contour (Ad-PAC) algorithm for the segmentation and tracking of the IJV in ultrasound videos. In the proposed algorithm, the parameters of the Ad-PAC algorithm are adapted based on the results of segmentation in previous frames. The Ad-PAC algorithm is applied to 65 ultrasound videos captured from 13 healthy subjects, with each video containing 450 frames. The results show that spatial and temporal adaptation of the energy function significantly improves segmentation performance when compared to current state-of-the-art active contour algorithms.
\end{abstract}


\section{Introduction}
\begin{figure}[t!]
\center
\includegraphics[width=0.75\linewidth]{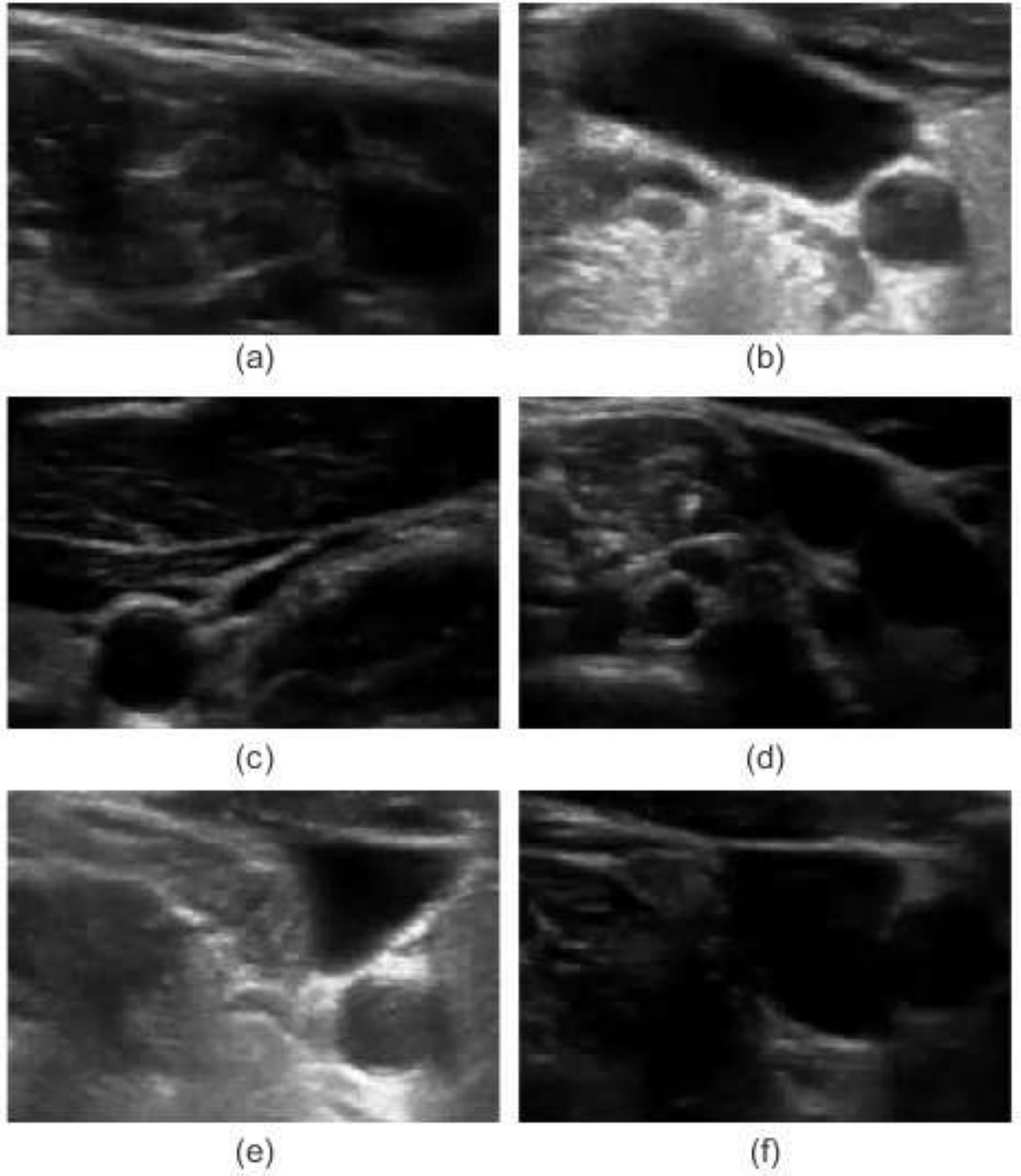} 
\caption{Sample images for IJV images with (a) low brightness, (b) high brightness, (c) fully collapsed, (d) partially missing contour (broken edge), (e) sharp contour (triangular) shape, and (f) non-convex contour shape.}\label{figims}
\end{figure}

\begin{figure}
\center
\includegraphics[width=0.9\linewidth]{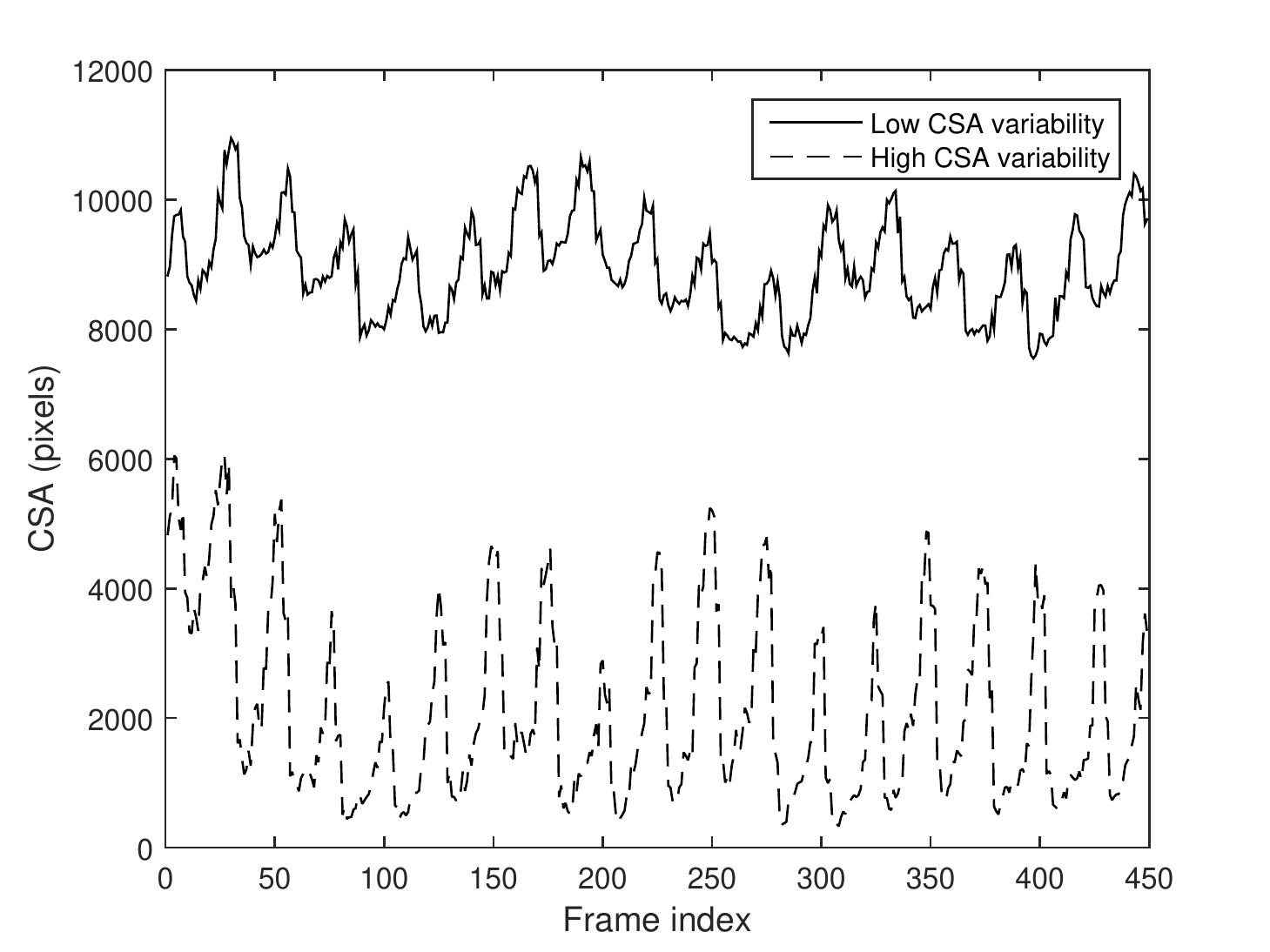} 
\caption{Temporal variations of IJV csa obtained from manual segmentation of two IJV videos with different CSA variability.}\label{figcsa}
\end{figure}

%\IEEEPARstart{S}
Determination of relative changes in circulating blood volume is important for a variety of acute and chronic medical conditions including hemorrhage from trauma, septic shock, dialysis and volume overload pertaining to congestive heart failure \cite{steuer1993,kudo1981,kasuya2003,yashiro2003,Bremer2004}. The estimation of absolute blood volume, while ideal, remains a significant challenge \cite{bose2006}. Recent studies suggest that non-invasive measures such as transverse ultrasound (cross-section area, CSA) of the internal jugular vein (IJV) can be used to detect and monitor relative changes in blood volume \cite{bailey2012,raksamani2014}. 
As shown in Fig \ref{figims}, the CSA of the IJV is dynamic with spatial and temporal variations that can correlate with relative changes in volume status. Short-term variability reflects a variety of factors including blood volume, proximity to the carotid artery, cardiac contractility, respiratory effort and local anatomy. Changes in parameters over the long-term can reflect relative changes in blood volume. Demonstration of short- and long-term CSA variability of a healthy patient sitting at different angles of inclination to simulate relative changes in circulating blood volume is shown in Fig. \ref{figcsa}. Accurate segmentation and tracking of the rapidly changing IJV is fundamental to the use of ultrasound to estimate relative changes in blood volume.

Portable ultrasound, the technology typically used to image the IJV in the acute care setting, does have its limitations. Ultrasound videos are generated at the point of care by clinicians with a broad range of skills which can result in significant variation in image quality. Furthermore, manual segmentation of IJV is a time-consuming task and inappropriate for real-time blood-volume monitoring applications.\\
\indent  Fully automatic segmentation algorithms are an ideal objective; however, they tend to require prior information about the images \cite{bala2015automatic,serra2012mathematical,diciotti2011}.
More recently, semi-automated segmentation algorithms requiring operator input have become popular in medical image processing \cite{zwiggelaar2003,xie2005,karasev2013interactive,spina2014hybrid}. For example, combinatorial graph-cut based algorithms perform the segmentation task by minimizing an energy function to find the minimal energy paths through the vertices that are manually selected by an operator \cite{price2010geodesic,grosgeorge2013graph,mahapatra2013automatic}. Unfortunately, graph-cut algorithms suffer from high computational complexity, making them inefficient for real-time frame-by-frame segmentation, and their results are sensitive to variations in image quality - a significant problem in ultrasound imaging.\\
\indent Similar to graph-cut algorithms, active contours (ACs) segment images via minimization of an energy function; however, their energy function is more flexible than graph-cut based algorithms. This is demonstrated in their ability to adapt to complex shapes and track temporal deformations, making them suitable for real-time monitoring applications in medicine. An additional advantage is that the minimization is performed over a continuous surface, improving the computational complexity over graph-cut based techniques \cite{meziou2012alpha,wang2014multiscale, yushkevich2006user, al2009active, paragios2000geodesic}. Unfortunately, these algorithms suffer from the fact that their performance is highly sensitive to the parameters of the energy function and the initial contour, resulting in a limited ability to track topological changes. This limitation was addressed in the geometric deformable AC models based on curve or surface evolution \cite{mesejo2015biomedical, ma2013segmentation, lee2012segmentation}. In these models, the evolution is independent of the parameter selection and topological variations, which makes them more suitable for object tracking; however, they are still unable to detect shape split or merge due to the low quality of ultrasound images.\\
\indent In general, the performance of AC algorithms also depends on the initial segmentation, and therefore, can be combined with other segmentation algorithms in a coarse-to-fine strategy. The coarse initial segmentation, obtained from another algorithm, provides a rough segmentation which is subsequently refined with an AC algorithm \cite{yim2003, ali2012}. In \cite{qian2014}, the combination of speckle tracking \cite{voigt2014definitions} and AC was proposed for the segmentation and tracking of the IJV in which the coarse segmentation obtained from speckle tracking was smoothed with an AC. Unfortunately, speckle tracking fails when the IJV undergoes fast variations, unless the ultrasound machine has a sufficiently high frame rate. This problem was addressed by cascading region growing \cite{wu2008} and AC (RGAC) \cite{karami2016}. Unfortunately, all of these methods continue to fail when the image quality is poor or when a part of the vessel wall is obscured by artifact.\\
\indent In the case of broken edges, active contours fail to resolve the contours of intersecting objects resulting in leakage. An active shape model (ASM) using a statistical shape model can be used to address the above mentioned problem \cite{ali2012integrated,sun2012automated}. Unfortunately, as per Fig. \ref{figims}, the IJV assumes many different shapes and therefore, ASM is not applicable for the IJV segmentation. Other common approaches, such as Kalman filers, have been proposed for real-time vessel tracking in ultrasound imagery; however, similar to ASM, they require the geometry of the vessel \cite{guerrero2007real}.
\par
\indent Segmentation can be viewed and solved as a spatio-temporal three dimensional (3D) segmentation problem with the time (frame index) defined as the 3rd dimension. 3D segmentation algorithms work on frame-by-frame basis using the similarity between regions  \cite{moscheni1998spatio,chen2013automatic,hamarneh2004deformable,karami2017a} or by attempting to minimize a 3D AC model \cite{li2010actin,mitchell2002,li2006optimal,ristivojevic2006space,smith2010segmentation}. The former methods again require imaging machines with high frame rates so that deformation from one frame to the next one is insignificant. The latter algorithms suffer from (1) significant computational complexity, (2) lower accuracy associated with minimization of a 3D energy function which involves more parameters, and (3) the entire video prior to initiation of segmentation eliminating the possibility of real-time monitoring.
\par
\indent Polar representation of the contour is a useful technique in a variety of medical image processing applications, particularly vessel segmentation, in which the shape of the object is generally convex \cite{zuo2004,collewet2009, baust2010, karami2017b}. Polar contours sample the object boundary at certain angles, reducing the degree of freedom for each contour point to one. In other words, the contour is evolved only radially, enabling the energy function to be minimized faster and more efficiently than conventional ACs. Several examples include \cite{zuo2004} in which polar edge detection is combined with AC for segmentation of tongue images; \cite{giannoglou2007} incorporated a polar active contour algorithm based on a classic energy function definition for segmentation of intra-vascular ultrasound images; \cite{collewet2009} used a polar active contour defined with the external force derived from an energy term based on the area inside the contour; and \cite{baust2011}, a variational polar active contour was proposed to inherit the robustness to local minima from Sobolev active contours. Unfortunately, all above mentioned polar AC algorithms are sensitive to image quality and object shape, with each working for a specific subset of image qualities and shapes. These algorithms fail to accurately segment clips across this spectrum, hence the need for an adaptive AC algorithm that accounts for these large variations.\\ 
\indent In this paper, a novel adaptive polar AC algorithm (Ad-PAC) is proposed for semi-automatic segmentation and tracking of the IJV videos. This algorithm involves the initial frame being manually segmented by an operator and subsequently serving as the reference for the initial energy function parameters selection. The parameters are then adapted from one frame to the next based on the segmentation results of previous frames. Section II introduces two related state-of-the-art polar AC algorithms; Section III describes the proposed algorithm; Section IV presents the results with comparisons to manual segmentation, traditional AC algorithms, and the two polar AC algorithms of Section II; and the conclusions are presented in Section V.
\section{Related Work}
\begin{figure}[t]
\centering
\includegraphics[width=1\linewidth]{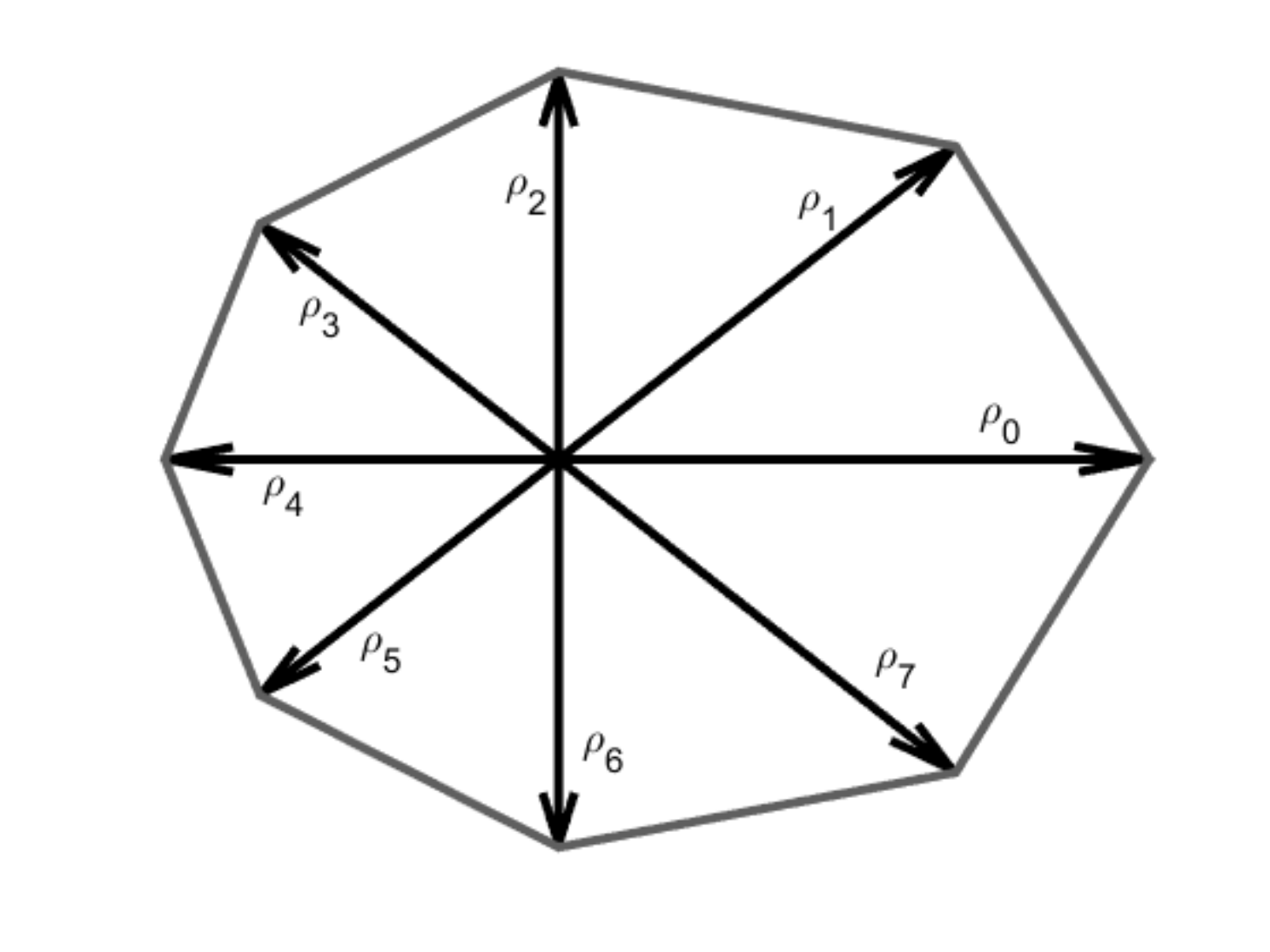} 
\vspace{-0.25in}
\caption{An example of polar contour with eight contour points.}\label{Polar}
\end{figure}

\noindent \textbf{Polar AC Algorithms}\\
\indent Polar representation of the contours reduce the degrees of freedom for each contour point such that the contour can only evolve radially. Each polar contour is sampled at certain angles, as shown in Fig. \ref{Polar} as a contour with $N=8$ points. 
%\textit{1) Polar Snake Algorithm \cite{de2014}:} 
In general, polar AC algorithms minimize the following energy function:
%\begin{equation}
%E_{total}=E_{int}+E_{ext},
%\end{equation}
\begin{equation}
E_{t}=E_{int}+E_{ext},
\end{equation}
where $E_{int}$ is the internal energy of the contour including forces from the contour shape parameters such as curvature ($E_{curv}$) and continuity ($E_{cont}$) defined as
\begin{equation}
E_{int}=\alpha E_{curv}+\beta E_{cont} ,
\end{equation}
\noindent where $\alpha$, $\beta$ are positive real constants. In \cite{de2014}, the curvature and continuity energies at angle $\theta$ are defined as

\begin{equation}
E_{curv,\theta}=(r_{\theta}-2r_{\theta+1}+r_{\theta+2})^2,\label{curvref}
\end{equation}

\begin{equation}
E_{cont,\theta}=(r_{\theta}-r_{\theta+1})^2,\label{contref}
\end{equation}

\noindent where $r_{\theta}$ is the radial distance at angle $\theta$. Furthermore, according to \cite{de2014}, $E_{ext,\theta}$ is the external Hibernian energy at angle $\theta$ and defined as
\begin{equation}
E_{ext,\theta}=\gamma \left(1-\frac{|\hat{f}(r_\theta)|}{max(|\hat{f}(r_\theta)|)}\right),
\end{equation}
\noindent where $\gamma$ is a positive real constant and $\hat{f}(r(\theta))$ is the Hilbert transform of $f(r(\theta))$.\\
%\textit{2) Variational Polar AC \cite{baust2012}:} \\In this algorithm, the energy function is defined as:
\indent To overcome the problem of having shape-dependent constants, variational polar AC models have been proposed in the past\cite{baust2010, baust2011}. In these models, the energy function has been defined as \cite{baust2011}:
\begin{equation}
E(c)=\int_{int(c)}f  dx +\alpha \int_{0}^{L}g  ds,
\end{equation}\noindent where $int(c)$ in the first integral that denotes the area which is bounded by the contour $c$, $L$ is the length of $c$, $f$ and $g$ indicate the intensity and gradient functions, respectively, and $\alpha$ is the weight given to boundary information. The $\mathbb{L}^2$-gradient flow of $E(c)$ is calculated as \cite{baust2011}:
\begin{equation}
\nabla_{\mathbb{L}^2} E(c)=Lf\textbf{n}+\alpha L(\nabla g \cdot \textbf{n}-g\kappa)\textbf{n},
\end{equation}
\noindent where $\bf{n}$ denotes the normal of the contour $c$ and $\kappa$ represents its curvature. 
\section{The Proposed Algorithm: Adaptive Polar AC (Ad-PAC)}
The proposed algorithm (Ad-PAC) can be categorized as a polar 2D AC where the parameters of the energy function are dynamically and automatically changed according to image quality as well as spatial and temporal variations of the object. Ad-PAC performs the segmentation task on a frame-by-frame basis creating the possibility for real-time applications. The novelties include:
\begin{itemize}
    \item Modified energy function: Ad-PAC enables researchers to incorporate many features followed by subsequent removal of weaker features depending on the application. An example of this is the continuity energy functional described in Section III-A - a weak feature for IJV segmentation which was subsequently removed. In this paper, we use six energy terms and derive their gradients in polar coordinates. Furthermore, we modified the energy terms proposed in existing polar ACs. For instance, from (\ref{curvref}) one can see that the curvature energy proposed in \cite{de2014} is not applicable, as in the case of a circular contour because it will result in an energy term equal to zero. This obviously cannot be correct as theoretically the curvature of a circle is the reciprocal of its radius and the curvature of a straight line is zero.
    \item Automatic adaptation of the energy function: Ad-PAC automatically adapts the parameters based on the spatial and temporal features of the object as follows:\\
    1. Spatial parameter adaptation: The parameters of the energy function are selected such that each energy term is optimized and adapted based on local features of objects such as shape and intensity. This makes the proposed algorithm more robust to image artifacts such as shadowing.\\ 
    2. Temporal parameter adaptation: In the proposed algorithm, the parameters of the energy function are adapted to temporal variations of the object, such as variations in shape, intensity, and the object area, on a per-frame basis. In conventional ACs, to achieve the best performance, parameters must be optimized for individual frames, a task requiring significant operator intervention. Except for the initialization, Ad-PAC automatically calculates optimized parameters drastically reducing the need for human intervention.
\end{itemize}
In the following subsections, we discuss these novelties in detail.
\subsection{The Energy Function}
In the proposed algorithm, we define the energy function as%\footnote{Note that since the proposed algorithm automatically and adaptively combines different energy terms, based on the information available from the object, the number of energy terms can be more than what we have introduced here.}
\begin{equation}
\begin{split}
&E=\alpha E_{curv}+\beta E_{cont}+\gamma E_{edge}\\&+\kappa E_{var}+\zeta E_{intensity}+\nu E_{contr},\label{energynew}
\end{split}
\end{equation}
where $\alpha$, $\beta$, $\gamma$, $\kappa$, $\zeta$, and $\nu$ are real positive numbers, and $E_{curv}$, $E_{cont}$, $E_{edge}$, $E_{var}$, $E_{intensity}$, and $E_{contr}$ are, respectively, the energy function corresponding to the information in the object curvature, continuity, boundary, variation of the intensity in and out of the contour, intensity on the object contour, and contraction energy.\\
$E_{curv}$ is the energy term used to control the contour curvature. From (\ref{curvref}), one can see that the curvature energy proposed in \cite{de2014} does not provide a correct result as it takes its minimum value for a circle, where $r_{\theta}=\mbox{constant}$, i.e., it cannot segment the boundaries with less curvature than a circle (e.g., a straight line). In this paper, we extend the curvature energy defined in the Cartesian coordinates, which is applicable to contours with any value of curvature, to polar coordinates as:
\begin{equation}
E_{curv}=\sum_{n=0}^{N-1}|p_{n+1}-2p_{n}+p_{n-1}|^2.\label{curvEn}
\end{equation}
\noindent where $N$ is the number of contour points, $|.|$ is the vector absolute value, and $p_{n}$ is the $n$th point vector defined as
\begin{equation}
\left[x_c+\rho_n \cos(n\phi_0),y_c+\rho_n \sin(n\phi_0)\right],\label{pointeq}
\end{equation}
\noindent with $x_c$ and $y_c$ being the coordinates of the center of the object in previous frame, respectively, $\phi_0=\frac{2\pi}{N}$. \\
$E_{cont}$ is the energy term used to control the distance between contour points and is defined as
\begin{equation} 
E_{cont}=\sum_{n=0}^{N-1}|p_{n+1}-p_{n}|^2.\label{DistEn}
\end{equation}
This definition is also similar to the continuity energy term used in Cartesian ACs and is different from the simple continuity energy term used in \cite{de2014}.\\
$E_{edge}$ represents the edge energy in the object boundary though this term has limited value in scenarios with indistinct edges as is often the case in ultrasound imaging. This energy term is defined as %is the energy in the object boundary. Although this energy term is not much useful in ultrasound imaging where object boundary is Fuzzy, but we still consider that not to limit the application of Ad-PAC for ultrasound imaging. 
\begin{equation}
E_{edge}=-\sum_{n=0}^{N-1}{|\nabla I\left(p_n\right)|^2},\label{GradEn}
\end{equation}
\noindent where $|\nabla I|$ is the gradient magnitude of the image and $I\left(p_n\right)$ is the image intensity at the current frame.\\
$E_{var}$ is similar to the variational energy term defined in either variational Cartesian or polar ACs and is defined as
\begin{equation}
E_{var}=-(u-v)^2,\label{VarEn}
\end{equation}
\noindent where $u$ and $v$ are the mean intensities inside and outside of the contour, respectively.\\
$E_{intensity}$ is a proposed energy term  which is defined to exploit the information in spatial illumination levels at the object boundary.
\begin{equation}
E_{intensity}=\sum_{n=0}^{N-1} |I\left(p_n\right)-I_0\left(p_n\right)|^2,\label{IntEn}
\end{equation}
\noindent where $I_0\left(p_n\right)$ is the reference intensity of the contour obtained from the previous segmented frame\\
$E_{contr}$ is the energy term that controls the area of the object and is defined as
\begin{equation}
E_{contr}=-\sum_{n=0}^{N-1} \rho_n, \label{ContrEn} 
\end{equation}
\subsection{Local Parameterization of the Energy Function}
\subsubsection{Overview}
\begin{figure}[t!]
\centering
\includegraphics[width=1\linewidth]{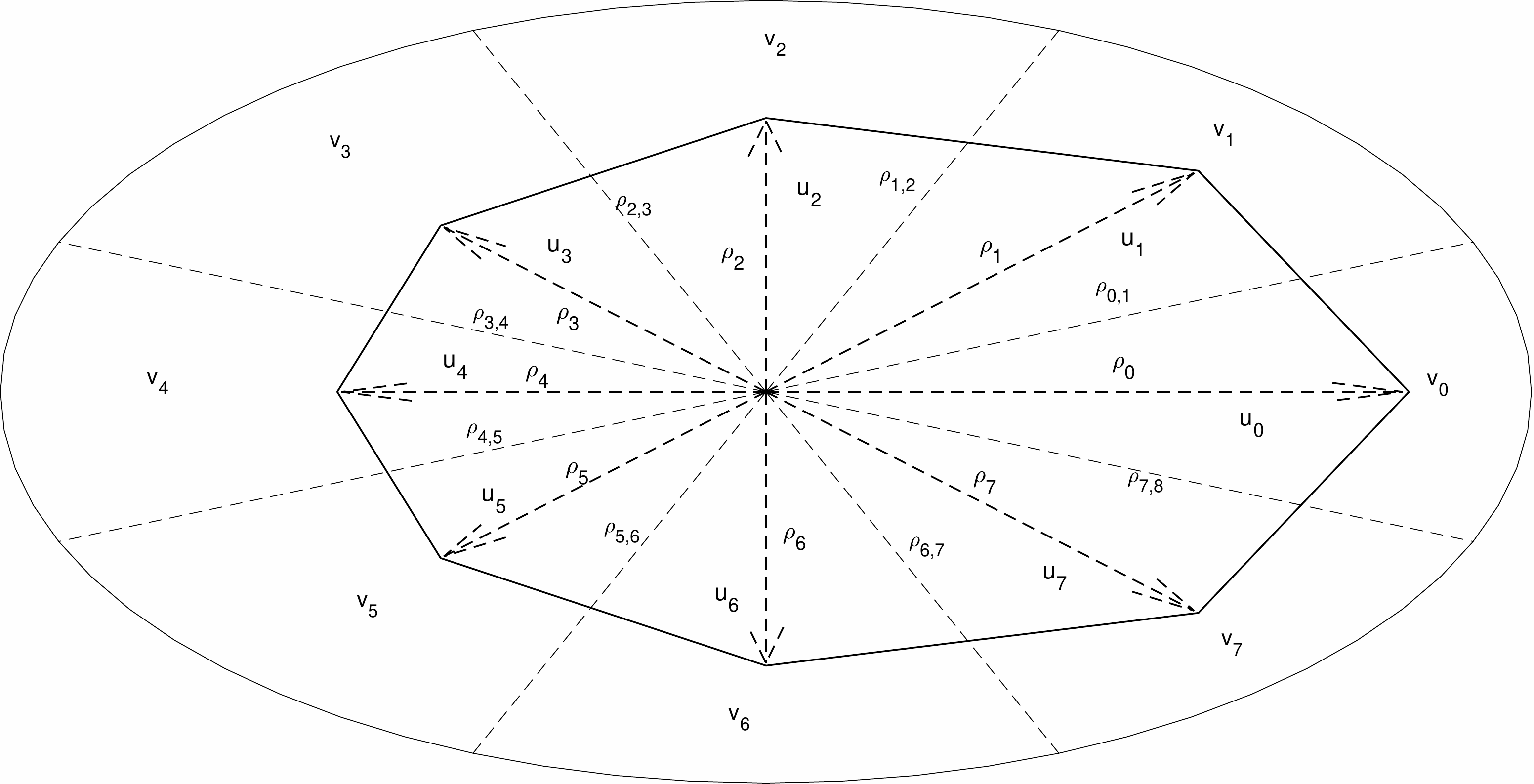} 
\caption{Adaptive polar structure.}\label{PolarSegs}
\end{figure}

A major shortcoming in existing AC algorithms is that they define similar energy terms for all contour points, despite the fact that intensity and shape vary across the region of interest. To overcome this problem, the proposed algorithm weights the energy terms locally. For example, if part of the IJV contour is obscured by shadow then the algorithm assigns smaller weights to the external energy of the points in the shadowed region such that increased emphasis is on the internal energy terms. Similarly, if a part of the contour has a sharp curvature, the algorithm gives a smaller weight to the curvature energy term for points in areas with sharp edges. These small, non-zero weights enable the contour points to have larger curvatures while still contributing to the total energy. Furthermore, regional variations in intensity are incorporated by subdividing the region of interest (ROI) into multiple sectors with each sector containing one contour point and values of $u$ and $v$ calculated locally as shown in Fig. (\ref{PolarSegs}).
After splitting the contour into $N$ sectors, the energy function is modified by giving spatial weights to equations (\ref{curvEn}-\ref{IntEn}) as:
\begin{equation}
E_{curv}=\sum_{n=0}^{N-1}\alpha_{n}|p_{n+1}-2p_{n}+p_{n+1}|^2,\label{curvloc}
\end{equation}
\begin{equation}
E_{cont}=\sum_{n=0}^{N-1}\beta_{n}|p_{n+1}-p_{n}|^2,\label{contloc}
\end{equation}
\begin{equation}
 E_{edge}=-\sum_{n=0}^{N-1}\gamma_{n} |\nabla I\left(p_{n}\right)|^2,\label{edgeloc}
\end{equation}
\begin{equation}
E_{var}=-\sum_{n=0}^{N-1}\kappa_{n} (u_{n}-v_{n})^2,\label{varloc}
\end{equation}
\begin{equation}
E_{intensity}=-\sum_{n=0}^{N-1}\zeta_{n}|I\left(p_n\right)-I_0\left(p_n\right)|^2,\label{intloc}
\end{equation}
\noindent where $\alpha_n$, $\beta_n$, $\gamma_n$, $\kappa_n$, and $\zeta_n$ are the weights given to the energy terms of the $n$th region. Note that $E_{contr}$ is the only remaining energy term which is not spatially adapted.
\subsubsection{Energy Functions in Polar Coordinates}
Polar representation is used to derive the energy function in the following sections.

\noindent The curvature energy: By substitution of (\ref{pointeq}) in (\ref{curvloc}), the curvature energy function is rewritten as
\begin{align}
E_{curv}=&\sum_{n=0}^{N-1}\alpha_{n}[4\rho_{n}^2+\rho_{n-1}^2+\rho_{n+1}^2-4\rho_{n}(\rho_{n-1}+\rho_{n+1})\nonumber\\&\times \cos(\phi_0)+2\rho_{n-1}\rho_{n+1}\cos(2\phi_0)],\label{curvPol}
\end{align}
\noindent and its gradient with respect to $\boldsymbol{\rho}$ is obtained as
\begin{equation}
\frac{\partial E_{curv}}{\partial \boldsymbol{\rho}}=\boldsymbol{A_c\rho},\label{GradEC}
\end{equation}
\noindent where $\textbf{A}_{c}$ is an $N \times N$ penta-diagonal matrix with its elements defined as
\begin{equation}
\begin{split}
&a_c(i,j)=\label{GradEC2}\\&\left\{
                \begin{array}{ll}
                  2\alpha_{i-1}\cos(2\phi_0), & \mbox{if} \quad mod(i-j,N)=2,\\
                  -4(\alpha_{i-1}+\alpha_{i})\cos(\phi_0), & \mbox{if} \quad mod(i-j,N)=1,\\
                  2(\alpha_{i-1}+4\alpha_{n}+\alpha_{n+1}), & \mbox{if} \quad i=j,\\
                  -4(\alpha_{i}+\alpha_{i+1}\cos(\phi_0), & \mbox{if} \quad mod(j-i,N)=1,\\
                  2\alpha_{i}\cos(2\phi_0), & \mbox{if} \quad mod(j-i,N)=2,\\
                  0, & \mbox{otherwise}.
                \end{array}
              \right.
\end{split}
\end{equation}
\noindent where $i,j=0,1,...,N-1$ are the indices of the matrix.\\\\
\noindent \textit{The continuity energy}: Similarly, by substitution of (\ref{pointeq}) in (\ref{contloc}), the continuity energy function is rewritten as
\begin{equation}
E_{cont}=\sum_{n=0}^{N}\left(\rho_{n}^2-2\rho_{n}\rho_{n+1}\cos(\phi_0)+\rho_{n+1}^2\right),\label{contPol}
\end{equation}
\noindent and its gradient with respect to $\boldsymbol{\rho}$ is computed as
\begin{equation}
\frac{\partial E_{cont}}{\partial \boldsymbol{\rho}}=\boldsymbol{B_c\rho},\label{GradED}
\end{equation}
\noindent where $\textbf{B}_{c}$ is an $N \times N$  tridiagonal matrix with its elements defined as
\begin{align}
b_c(i,j)=&\label{GradED2}\\&\left\{                
\begin{array}{ll}
                  -2\beta_{i-1}\cos(\phi_0), & \mbox{if} \quad mod(i-j,N)=1,\\
                  2(\beta_{i}+\beta_{i-1}), & \mbox{if} \quad i=j,\\
                  -2\beta_{i}\cos(\phi_0), & \mbox{if} \quad mod(j-i,N)=1,\\                 
                  0, & \mbox{otherwise}.
                \end{array}
              \right. \nonumber
\end{align} 
\\
\noindent \textit{The edge energy}: From (\ref{edgeloc}), the gradient of the edge energy with respect to $\boldsymbol{\rho}$ is an $N \times 1$ vector $\boldsymbol{G}=\left[ g_0,\quad g_1,\quad ..., \quad g_{N-1}\right]^{T}$ where
\begin{align}
g_i=&\label{GradEE}\\&-\gamma_{i} \left[\frac{\partial  |\nabla I\left(p_{i}\right)|^2}{\partial x}\cos(i\phi_0)+\frac{\partial  |\nabla I\left(p_{i}\right)|^2}{\partial y}\sin(i\phi_0)\right].\nonumber
\end{align}
\noindent \textit{The variational energy}: In (\ref{varloc}), $u_{n}$ and $v_{n}$ can be computed as,
\begin{equation}
u_n=\frac{S_n}{A_n},\label{un}
\end{equation}
\begin{equation}
v_n=\frac{S_n^{(c)}}{A_n^{(c)}},\label{vn}
\end{equation}
\noindent where $S_n$ and $S_n^{(c)}$ are total intensities inside $C_n$ and $C_n^{(c)}$, respectively, and are obtained as
\begin{equation}
S_n=\int_{C_n}  I dA,
\end{equation}
\begin{equation}
S_n^{(c)}=\int_{C_n^{(c)}}  I dA,
\end{equation}
\noindent with $C_n$ and $C_n^{(c)}$ as the areas inside and outside the $n$th sector, respectively, and $A_n$ and $A_n^{(c)}$ are their corresponding areas which can be computed as follows,
\begin{equation}
A_n\approx \frac{1}{4}\sin\left(\frac{\phi_0}{2}\right)\rho_n\left(\rho_{n+1}+2\rho_{n}+\rho_{n-1}\right), \label{sectAr1}
\end{equation}
\begin{equation}
A_n^{(c)}=\frac{1}{2}\phi_0 R^2-A_n, \label{sectAr2}
\end{equation}
\noindent where $R$ is the maximum radius of the search area. In this paper, $R$ is set to be 50 percent larger than the maximum contour radius, $\rho_{n}$, estimated in previous frame. \\
\indent From (\ref{un}) and (\ref{vn}), the gradient of $u_n$ and $v_n$ with respect to $\rho_i$ is computed as,
\begin{equation}
\frac{\partial u_n}{\partial \rho_i}=\frac{\left(I_n-u_n\right)}{A_n}\frac{\partial A_n}{\partial \rho_i}, \label{GradU}
\end{equation}
\begin{equation}
\frac{\partial v_n}{\partial \rho_i}=-\frac{\left(I_n-v_n\right)}{A_n}\frac{\partial A_n}{\partial \rho_i}, \label{GradV}
\end{equation}
\indent From (\ref{VarEn}), (\ref{GradU}), (\ref{GradV}), (\ref{sectAr1}), and (\ref{sectAr2}), one can find the gradient of $E_{intensity}$ with respect to $\rho_n$ as
\begin{equation}
\frac{\partial E_{var}}{\partial \rho}=\boldsymbol{U\rho},\label{GradEI}
\end{equation} 
\noindent where $\boldsymbol{U}$ is an $N \times N$ bi-diagonal matrix defined as follows,
\begin{equation}
U(i,j)=\left\{\begin{array}{ll}
\iota(i)+\iota(i-1) & \mbox{if} \quad mod(i-j,N)=1,\\
4\iota(i) & \mbox{if} \quad i=j,\\
\iota(i)+\iota(i+1) & \mbox{if} \quad mod(j-1,N)=1,\\
                  0, & \mbox{otherwise},
\end{array}
\right.\label{GradEI2}
\end{equation}
\noindent with $\iota(i)$ defined as
\begin{align}
\iota(i)=&\\&-\frac{1}{2}\kappa_i\sin(\frac{\phi_0}{2})(u_i-v_i)\left(\frac{I_i-u_i}{A_i}-\frac{I_i-v_i}{\frac{1}{2}\phi_0 R^2-A_i}\right).\nonumber
\end{align}
\noindent \textit{The intensity energy:} the gradient of the intensity energy with respect to $\boldsymbol{\rho}$ is an $N \times 1$ vector $\boldsymbol{\chi}$, in which its elements are defined as
\begin{align}
\boldsymbol{\chi}_i=&\zeta_i\left(I\left(p_i\right)-I_0\left(p_i\right)\right)\label{GradEIN}\\&\times \left(\frac{\partial I\left(p_{i}\right)}{\partial x}\cos(i\phi_0)+\frac{\partial  I\left(p_{i}\right)}{\partial y}\sin(i\phi_0)\right).\nonumber
\end{align}
\indent In this paper, simple iterative gradient descent technique is used to minimize the defined energy function, although due to single-dimentionality of the energy function, faster techniques are also applicable \cite{jiang2013region}.  From equations (\ref{GradEC}), (\ref{GradED}), (\ref{GradEE}), (\ref{GradEI}), and (\ref{GradEIN}) the gradient of the total energy function is obtained as
\begin{equation}
\frac{\partial E_c}{\partial \boldsymbol{\rho}}=\left(\boldsymbol{\alpha A_c+\beta B_c+\kappa U}\right)\boldsymbol{\rho}+\boldsymbol{\gamma G+\zeta \chi}+\nu\boldsymbol{1_{N \times 1}}.\label{GradET}
\end{equation}
\noindent where $\boldsymbol{1_{N \times 1}}$ is $N \times 1$ all-ones vector. Using (\ref{GradET}), the energy function can be iteratively minimized as
\begin{align}
\boldsymbol{\rho}_{(i)}=&\boldsymbol{\rho}_{(i-1)}\label{itmin}-\mu_1(\boldsymbol{1}+\mu_2\boldsymbol{\rho}_{(i-1)})\\&\odot \left[\left(\boldsymbol{\alpha A_c+\beta B_c+\kappa U}\right)\boldsymbol{\rho}+\boldsymbol{\gamma G+\zeta \chi}+\nu\boldsymbol{1_{N \times 1}}\right],\nonumber
\end{align}
\noindent where $\boldsymbol{\rho}_{(i)}$ is the estimation of $\boldsymbol{\rho}$ at the $k$th iteration, $\odot$ is the Hadamard vector product, and $\mu_1$ and $\mu_2$ are two step size parameters that let the algorithm to converge faster to the equilibrium when the contour point is farther from the center point.

\subsection{Parameter Adaptation}
The parameters are adapted to the shape and size of the object and to the local intensities around the contour points of previous frames.

\noindent \textit{Adaptation of the number of contour points $N$}: The first parameter adapted is the number of contour points, $N$, whose optimal value depends on the object size. Adaptation of $N$ is important due to large size variations in the contour. For example, when the contour shrinks, the elements of $\boldsymbol{\rho}_{(i)}$ become very small such that some terms can become negative during energy minimization using (\ref{itmin}). On the contrary, when the contour expands, the distance between adjacent contour points increases resulting in an inaccurate rough contour. In this paper, we select the number of contour points such that the average spacing between contour points is a constant value $\Lambda$ while later demonstrating the influence of $\Lambda$ on the segmentation accuracy. After segmenting each frame, the perimeter  of the contour ($P$) is computed and used to calculate the number of contour points using
\begin{equation}
N^{k}=\left\lceil{\frac{\sum_{n=0}^{N^{k-1}-1}\left|p_{n+1}^{k-1}-p_n^{k-1}\right|}{\Lambda}}\right\rceil, \label{upN}
\end{equation}
\noindent where $\left \lceil{x}\right \rceil $ denotes the smallest integer greater than or equal to $x$, and superscript $k$ denotes the frame index. After updating the number of contour points, the contour is re-sampled at angles 
\begin{equation}
\phi_n=n\phi_0^k,\label{upangles}
\end{equation}
\noindent where $n=0,1,...,N^{k}-1$ and $\phi_0^k=\frac{2\pi}{N^k}$.\\
\textit{Parameter selection for the curvature energy}: If the object boundary around a contour point is sharp, then the gradient of the curvature energy of that contour point is relaxed such that at the equilibrium (total energy minimized) it can approach a value with larger absolute value compared to the points on more smooth parts of the contour. Therefore, in the proposed algorithm, we select the parameters $\alpha_n^{k}$ of the $k$th frame, such that at the equilibrium condition for the reference frame (previous frame), all contour points have similar gradient of curvature as
\begin{equation}
 \left|\frac{\partial E_{curv,k,k-1}}{\partial \rho_n^{k-1}}\right|=1,\label{alfin}
\end{equation}
\noindent where $E_{curv,k,k-1}$ is defined as the curvature energy of the $(k-1)$th frame with the weights $\alpha_n^{k}$ and is computed by substituting $\alpha_n=\alpha_n^{k}$ and $\rho_n=\rho_n^{k-1}$ in (\ref{curvPol}). Using (\ref{GradEC}) and (\ref{GradEC2}), (\ref{alfin}) can be rewritten as 
\begin{equation}
\left|\boldsymbol{A_c}\boldsymbol{\rho}^{k-1}\right|=\boldsymbol{1_{N \times 1}}.\label{curv1st}
\end{equation}
\noindent where the elements of $\boldsymbol{A_c}^k$ are calculated using equation ($\ref{GradEC2}$) for the $k$th frame. Note that equation (\ref{curv1st}) is NP-hard, but we can find an approximate closed form solution as follows. The $n$th row of the matrix equation in (\ref{curv1st}) is
\begin{align}
|&2\alpha_{n-2}\rho_{n-2}^{k-1}\cos(2\phi_0)-4(\alpha_{n-1}+\alpha_{n})\rho_{n-1}^{k-1}\cos(\phi_0)\nonumber\\&+2(\alpha_{n-1}+4\alpha_{n}+\alpha_{n+1})\rho_{n}^{k-1}-4(\alpha_{n}\nonumber\\&+\alpha_{n+1})\rho_{n+1}^{k-1}\cos(\phi_0)+2\alpha_{n+2}\rho_{n+2}^{k-1}\cos(2\phi_0)|\nonumber\\&=1.\label{curv2nd}
\end{align}
By assuming that the local parameters $\alpha_n$ do not rapidly change, i.e., $\alpha_{n-2}\approx\alpha_{n-1}\approx\alpha_{n}\approx\alpha_{n+1}\approx\alpha_{n+2}$, (\ref{curv2nd}) is simply as
\begin{align}
\alpha_n^{k}=&\left(\epsilon+\left|12\boldsymbol{\rho_n}^{k-1}-8(\boldsymbol{\rho_{n-1}}^{k-1}+\boldsymbol{\rho_{n+1}}^{k-1})\cos(\phi_0^{k})\right.\right.\nonumber\\&\left.\left.+2(\boldsymbol{\rho_{n_{-2}}}^{k-1}+\boldsymbol{\rho_{n_{+2}}}^{k-1})\cos(2\phi_0^{k})\right|\right)^{-1},\label{alflast}
\end{align}
where $\epsilon$ is a very small number to avoid zero at the denominator. Note that with this parameter selection, (\ref{alfin}) is approximately satisfied.

\noindent \textit{Parameter selection for the continuity energy}: Similarly, the weights of continuity energy are adapted using the segmentation result from the previous frame such that
\begin{equation}
 \left|\frac{\partial E_{cont,k,k-1}}{\partial \rho_n^{k-1}}\right|=1,\label{betin}
\end{equation}
\noindent where $E_{cont,k,k-1}$ is the continuity energy of the $(k-1)$th frame with the weights $\beta_n^{k}$ and is computed by substituting $\beta_n=\beta_n^{k}$ and $\rho_n=\rho_n^{k-1}$ in (\ref{contPol}). Using (\ref{GradED}) and (\ref{GradED2}), (\ref{betin}) can be rewritten as 
\begin{equation}
\left|\boldsymbol{B_c}^k\boldsymbol{\rho}^{k-1}\right|=\boldsymbol{1_{N \times 1}},\label{cont1st}
\end{equation}
\indent Similar to (\ref{curv1st}), (\ref{cont1st}) is NP-Hard but has an approximate solution as
\begin{align}
\beta_n^{k}=&\left(\epsilon+|4\boldsymbol{\rho_n}^{k-1}-2(\boldsymbol{\rho_{n-1}}^{k-1}+\boldsymbol{\rho_{n+1}}^{k-1})\cos(\phi_0^{k})|\right)^{-1}.\label{betlast}
\end{align}
\noindent \textit{Parameter selection for the edge energy}: Adaptation of the parameters $\gamma_n$ in the edge energy function (see eq. \ref{GradEn})) is crucial. Lack of adaption would result in the edge energy forcing the AC to continue to expand in areas of broken edges or shadow resulting in leakage outside the actual object boundary. The weighting for the edge energy is selected such that
\begin{equation}
 \left|\frac{\partial E_{edge,k,k-1}}{\partial \rho_n^{k-1}}\right|=1,\label{gamin}
\end{equation}
\noindent where $E_{edge,k,k-1}$ is $|\nabla I_{1}\left(p_{n}\right)|^2$ of the $(k-1)$th frame. Consequently, the weights $\gamma_n^{k}$ are obtained as
\begin{align}
\gamma_n^k=&\left(\epsilon+\left|\frac{\partial  |\nabla I_{k-1}\left(p_{n}^{k-1}\right)|^2}{\partial x}\cos(n\phi_0^{k})\right.\right.\nonumber\\&\left.\left.+\frac{\partial  |\nabla I_{k-1}\left(p_{n}^{k-1}\right)|^2}{\partial y}\sin(n\phi_0^{k})\right|\right)^{-1}.\label{gamlast}
\end{align}
\noindent \textit{Parameter selection for the variational energy}: Similar to the previous energy terms, the weights $\kappa_n$ in the variational energy function (see equation (\ref{VarEn})) are selected such that
\begin{equation}
 \left|\frac{\partial E_{var,k,k-1}}{\partial \rho_n^{k-1}}\right|=1,\label{kapin}
\end{equation}
\noindent where $\kappa$ is a constant value and $E_{var,k,k-1}$ is the variational energy of the $(k-1)$th frame with the weights $\kappa_n^{k}$ and is computed by substituting $\kappa_n=\kappa_n^{k}$ and $\rho_n=\rho_n^{k-1}$ in (\ref{VarEn}). Consequently, using (\ref{GradEI}), (\ref{GradEI2}), and (\ref{kapin}), the weights $\kappa_n^k$ can be approximately obtained as
\begin{align}
\kappa_n^{k}=&\left(\epsilon+\left|\left(\iota^{k-1}(n)+\iota(n-1)\right)\rho_{n-1}^{k-1}+4\iota^{k-1}(n)\rho_n^{k-1}\right.\right.\nonumber\\&\left.\left.+\left(\iota^{k-1}(n)+\iota^{k-1}(n+1)\right)\rho_{n+1}^{k-1}\right|\right)^{-1}.\label{kaplast}
\end{align}
\noindent \textit{Parameter selection for the intensity energy:} The intensity energy and its gradient at the previous frame are both equal to zero (see eq. (\ref{IntEn})) and hence, the weights cannot be based on the information from previous frame. In this paper, the weights $\zeta_n^{k}$ are heuristically set as
\begin{equation}
\zeta_n^{k}=\left|\left(I\left(p_i^{k-1}\right)\right)\right|^2,\label{zetlast}
\end{equation}
From (\ref{zetlast}), one can see that in the case of broken edge where the intensity is almost zero, the intensity energy is automatically set to zero which is a reasonable outcome.\\
\noindent \textit{Parameter Adaptation with Forgetting Factor}:
Since the parameters of the AC are not expected to change rapidly, the parameters obtained from previous frames can always be used to improve segmentation accuracy and avoid misleading results due to the poor segmentation of any one individual frame. The forgetting factor, $\xi$, is defined as
\begin{eqnarray}
\alpha_n^k=\xi\alpha_{n,0}^k+(1-\xi)\alpha_n^{k-1},\label{alff}\\
\beta_n^k=\xi\beta_{n,0}^k+(1-\xi)\beta_n^{k-1},\label{betf}\\
\gamma_n^k=\xi\gamma_{n,0}^k+(1-\xi)\gamma_n^{k-1},\label{gamf}\\
\kappa_n^k=\xi\kappa_{n,0}^k+(1-\xi)\kappa_n^{k-1},\label{kapf}\\
\zeta_n^k=\xi\zeta_{n,0}^k+(1-\xi)\zeta_n^{k-1},\label{zetf}
\end{eqnarray}
\noindent where $\alpha_{n,0}^k$, $\beta_{n,0}^k$, $\gamma_{n,0}^k$, and $\kappa_{n,0}^k$ are the value of the parameters obtained from (\ref{alflast}), (\ref{betlast}), (\ref{gamlast}), and (\ref{kaplast}), respectively. Note that since the first frame is segmented manually, its results are assumed to be accurate and therefore, the forgetting factor $\xi$ is applied to weights obtained from the segmentation results of subsequent frames. Also note that, since it is assumed that at equilibrium, the overall gradient of energy is zero, then
\begin{equation}
\alpha+\beta \approx \gamma+\kappa.\label{equillib}
\end{equation} 
\section{Implementation of The Ad-PAC Algorithm}
The Ad-PAC algorithm is initialized from a manually generated segmentation of the first frame. This manual segmentation is adjusted and smoothed using Ad-PAC (without parameter adaptation), reducing the error associated with initial manual segmentation. This makes the initial segmentation insensitive to small operator errors as large as 5 pixels. For larger operator errors, this smoothing process may shift the initial manual segmentation to a different local minima (such as the boundary of an adjacent object). Next, the number of contour points is updated using (\ref{upN}). Third, the centroid is calculated as
\begin{eqnarray}
x_c^{k}=x_c^{k-1}+\frac{1}{N^{k}}\sum_{n=0}^{N^{k-1}-1}\rho_n^{k-1}\cos(\phi_n^k),\label{upxc}\\
y_c^{k}=y_c^{k-1}+\frac{1}{N^{k}}\sum_{n=0}^{N^{k-1}-1}\rho_n^{k-1}\sin(\phi_n^k)\label{upyc},
\end{eqnarray}
followed by the contour being re-sampled at angles obtained from (\ref{upangles}) using a cubic spline interpolation algorithm\cite{fritsch1980monotone}. Next, the weights of the energy function are obtained using  (\ref{alflast}), (\ref{betlast}), (\ref{gamlast}), (\ref{kaplast}), and (\ref{zetlast}). Then, the value of $\boldsymbol{\rho}^{(k)}$ is iteratively updated using (\ref{itmin}), until the equilibrium stop condition is met. In this paper, the algorithm is assumed to be at equilibrium if $\max_i|\rho^{(k)}_i-\rho^{(k-1)}_i|<10^{-4}$, where $\max(.)$ is element-wise maximum.
Finally, the algorithm returns to the first step to repeat the procedure for the next frame. A summary of the Ad-PAC algorithm is shown in table 1.

\floatname{algorithm}{}
\begin{algorithm}

\renewcommand{\thealgorithm}{}
\caption{\textbf{Table 1. Ad-PAC algorithm}}
\begin{algorithmic}[0]
%\STATE \small{\textbf{Required signal pre-processing:} Blind carrier frequency and timing synchronization, blind identification of the modulation type, blind estimation of the channel coefficients, and estimation of the noise power.}
\STATE \textbf{Input:} A video, $p_{n}^{(1)}$, with $n=0,1,...,N-1$ as manual segmentation of the first frame, where $N$ is the initial number of contour points, and parameters $\alpha$, $\beta$, $\gamma$, $\kappa$, $\zeta$, and $\Lambda$.
\STATE - Read one frame from the input video.
\STATE - Update the number of contour points using (\ref{upN}).
\STATE - Find the center of the contour and re-sample the contour at the angles $\phi=\frac{2n\pi}{N}$.
\STATE - Update the weights of the energy function using (\ref{alflast}), (\ref{betlast}), (\ref{gamlast}), and (\ref{kaplast}), (\ref{zetlast}), (\ref{alff}), (\ref{betf}), (\ref{gamf}), and (\ref{kapf}), and (\ref{zetf}).
\STATE - Update the value of $\boldsymbol{\rho}$ iteratively, using (\ref{itmin}).
\STATE - Repeat the previous step until the algorithm reaches to the equilibrium condition. The equilibrium condition is defined as the condition when the maximum absolute value of change in $\boldsymbol{\rho}$ at the previous step is less than $10^{-4}$ pixels.
\STATE - Return to the first step for the next frame.
\end{algorithmic}
\end{algorithm}
\vspace{.5cm}
\textit{Computational complexity:} The computational complexity of the Ad-PAC algorithm is estimated using the number of floating point operations (flops) \cite{watkins2004fundamentals}. From \ref{GradEC}, \ref{GradEC2}, \ref{GradED}, \ref{GradED2}, \ref{GradEE}, \ref{GradEI}, \ref{GradEI2}, \ref{GradEIN} and \ref{itmin}, one can see that the total number of flops required for each iteration of the algorithm is $70N$. From (\ref{alflast}), (\ref{betlast}), (\ref{gamlast}), and (\ref{kaplast}), (\ref{zetlast}), (\ref{alff}), (\ref{betf}), (\ref{gamf}), and (\ref{kapf}), and (\ref{zetf}), one can further see that the total number of flops required to parameter adaptation is $61N$ plus an additional $30N$ flops required for cubic re-sampling \cite{meijering2003note}, and $10N+2$ flops to update the center of the contour. Additionally, to compute the Sobel gradients of a 380 by 365 frame, $N_{G}=832200$ flops are required. Consequently, assuming $N_{iter}$ as the number of iterations required to minimize the energy function, the total number of $N_{flops}\approx 70N_{iter}N+101N+N_G+2$ is required for segmentation of each frame with Ad-PAC algorithm. Note that since parameter adaption is performed only once per frame, it does not significantly affect the processing time. Assuming $N_{iter}=5000$, $N=64$, and 25 percent extra processing power required for the software overhead,  with an average Intel Core i750 the segmentation of each 380 by 365 frame requires only $3\times10^{-3}$ sec, which makes the algorithm suitable for real time segmentation and tracking purpose. Note that the running time can be significantly reduced by using faster minimization techniques such as the dynamic programming approach described in \cite{jiang2013region}.
%\newpage
%\renewcommand{\bibname}{References}
\section{Results}
The experimental data was collected from 13 healthy subjects and with head of the bed elevated at 0, 30, 45, 60, and 90 degrees to simulate relative changes in blood volume. The IJV was imaged in the transverse plane using a portable ultrasound (M-Turbe, Sonosite-FujiFilm) with a linear-array probe (6-15 Mhz). Each video has a frame rate of 30 fps, scan depth of 4cm, and a duration of 15 seconds (450 frames/clip). The study protocol was reviewed and approved by the Health Research Ethics Authority.
\par Ad-PAC performance was compared to expert manual segmentation, Ad-PAC without parameters adaptation, Ad-PAC without temporal adaptation, and two current state-of-the-art polar AC algorithms introduced in Section II \cite{baust2011,de2014}. Additionally, it is also compared to region growing (RG) \cite{wu2008} and its combination with AC (RGAC) \cite{karami2016}, Speckle tracking driven AC (STAC) \cite{qian2014}, and two classic AC algorithms - Chan-Vese \cite{chan2001} and Geodesic \cite{caselles1997}. For each of these algorithms, parameter optimization was accomplished using a small subset of videos with variable image quality. For each video, the first frame was manually segmented by an operator with subsequent frames segmented automatically. \par
Initial efforts involved noise filtration using a variety of median and bilateral filters; however, no performance improvement was noted on any of the algorithms. This is mainly due to the fact that speckle noise includes useful information as it is random but deterministic that can improve the performance of AC algorithms. Hence, pre-processing techniques were not employed throughout this research.
\par
\indent The average contour point spacing was defined to be 10 pixels for the Ad-PAC and the other four algorithms $N=64$. Image intensities were normalized to between 0 and 1. The parameters of Ad-PAC algorithm were empirically set to be $\alpha=1$, $\beta=0$, $\gamma=0.05$, $\kappa=0.8$, $\zeta=150$, $\nu=0.0012$, $\mu_1=10^{-4}$, $\mu_2=1$, and $\epsilon=10^{-4}$. After segmentation of each frame, the maximum range $R$ was readjusted to be $1.5 \rho_{max}$, where $\rho_{max}$ was the largest element of $\boldsymbol{\rho}$ obtained from the previous frame.
\subsection{Evaluation of Extraction}
Before we define the validation metrics, we need to define the following terms:\\
\textit{True positive (TP)}: The number of pixels correctly segmented as foreground (manual
segmentation overlapping with algorithm segmentation) is True Positive (TP) and defined as
\begin{equation}
    TP=|\mathbb{A}\cap \mathbb{M}|,\label{eq_tp}
\end{equation}
\noindent where $\mathbb{A}$ and $\mathbb{M}$ are the set of pixels inside the contour obtained from the algorithms and manual segmentation, respectively, $|\mathbb{A}\cap \mathbb{M}|$ the intersection of the area between them, and $|.|$ denotes the cardinality of the set. \\
\textit{False positive (FP)}: False positive (FP) is the number
of pixels that are falsely segmented as foreground and is represented as
\begin{equation}
    FP=|\mathbb{A}\cap \mathbb{M}^c|,\label{eq_fp}
\end{equation}
\noindent where superscript $^c$ denotes set complement, i.e., set of the pixels outside the contour.\\
\textit{True negative (TN)}: True negative (TN) is the number of
pixels correctly labeled as background and defined as
\begin{equation}
    TN=|\mathbb{A}^c\cap \mathbb{M}^c|,\label{eq_tn}
\end{equation}
False negative (FN) is the number of pixels falsely detected
as background and defined as
\begin{equation}
    FN=|\mathbb{A}^c\cap \mathbb{M}|,\label{eq_fn}
\end{equation}
Using these terms, \textit{sensitivity} and \textit{specificity} which are also known as true positive and false positive rates, respectively, are obtained as
\begin{equation}
    Sensitivity=\frac{TP}{TP+FN},
\end{equation}
\begin{equation}
    Specificity=\frac{FP}{FP+TN}.
\end{equation}
The DICE factor (DF), also known as Sorensen, is the most common metric used to determine the correlation between algorithm and manual segmentation results \cite{dice1945measures}. The DICE coefficient, $DF$, is defined as:
\begin{equation}
DF=\frac{2|\mathbb{A}\cap \mathbb{M}|}{|\mathbb{A}|+|\mathbb{M}|},
\end{equation} 
\noindent and can be obtained from the above metric as
\begin{equation}
DF=\frac{2TP}{2TP+FP+FN}.
\end{equation}

\subsection{Influence of Initial Parameter Selection on the Performance of Ad-PAC}\label{influence}
\begin{figure}[tb!]
\center
\includegraphics[width=0.9\linewidth]{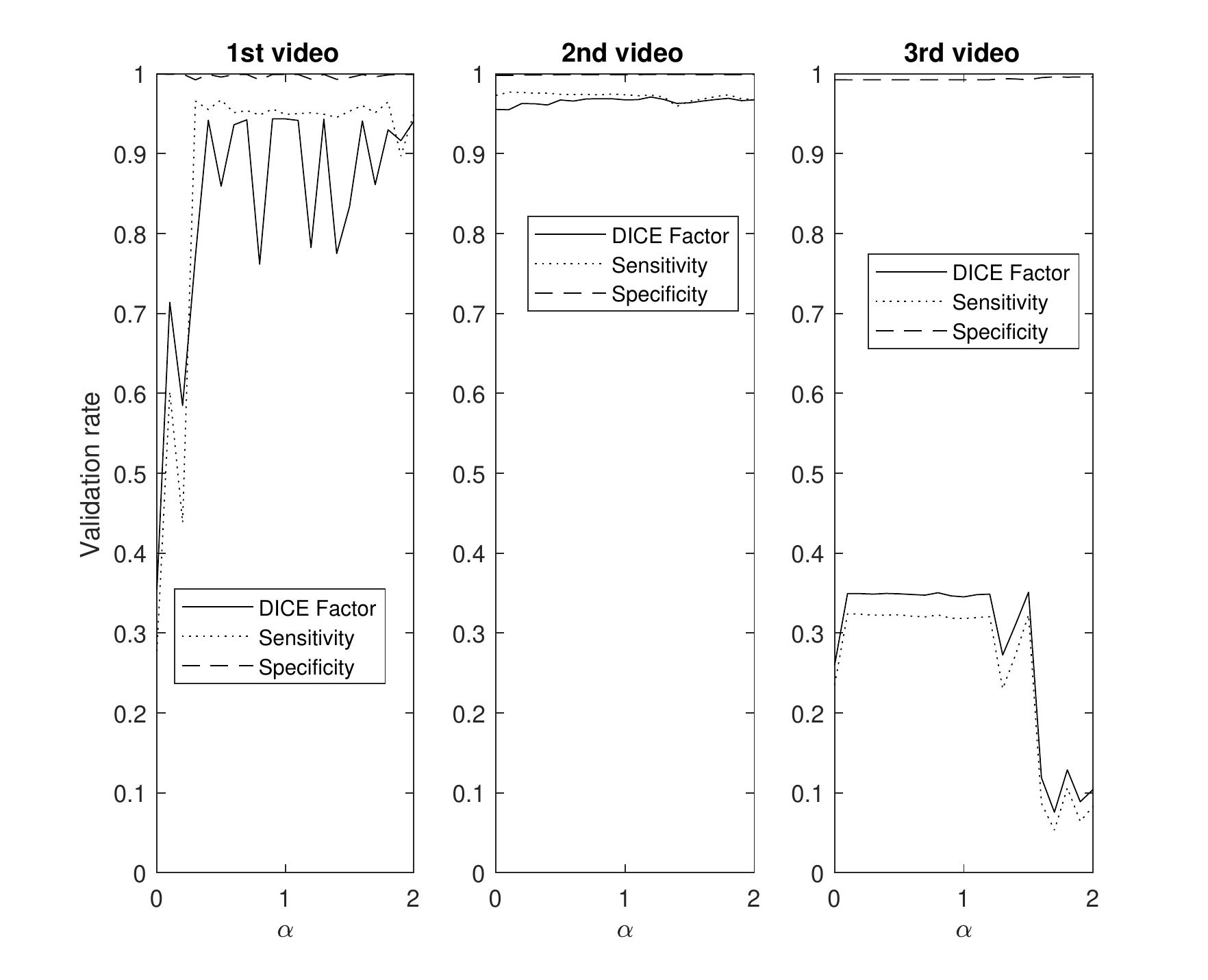} 
\vspace{-0.5cm}
\caption{The validation rates in terms of DF, sensitivity, and specificity versus the parameter $\alpha$.}\label{Ralpha}
\end{figure}

\begin{figure}[tb!]
\center
\includegraphics[width=0.9\linewidth]{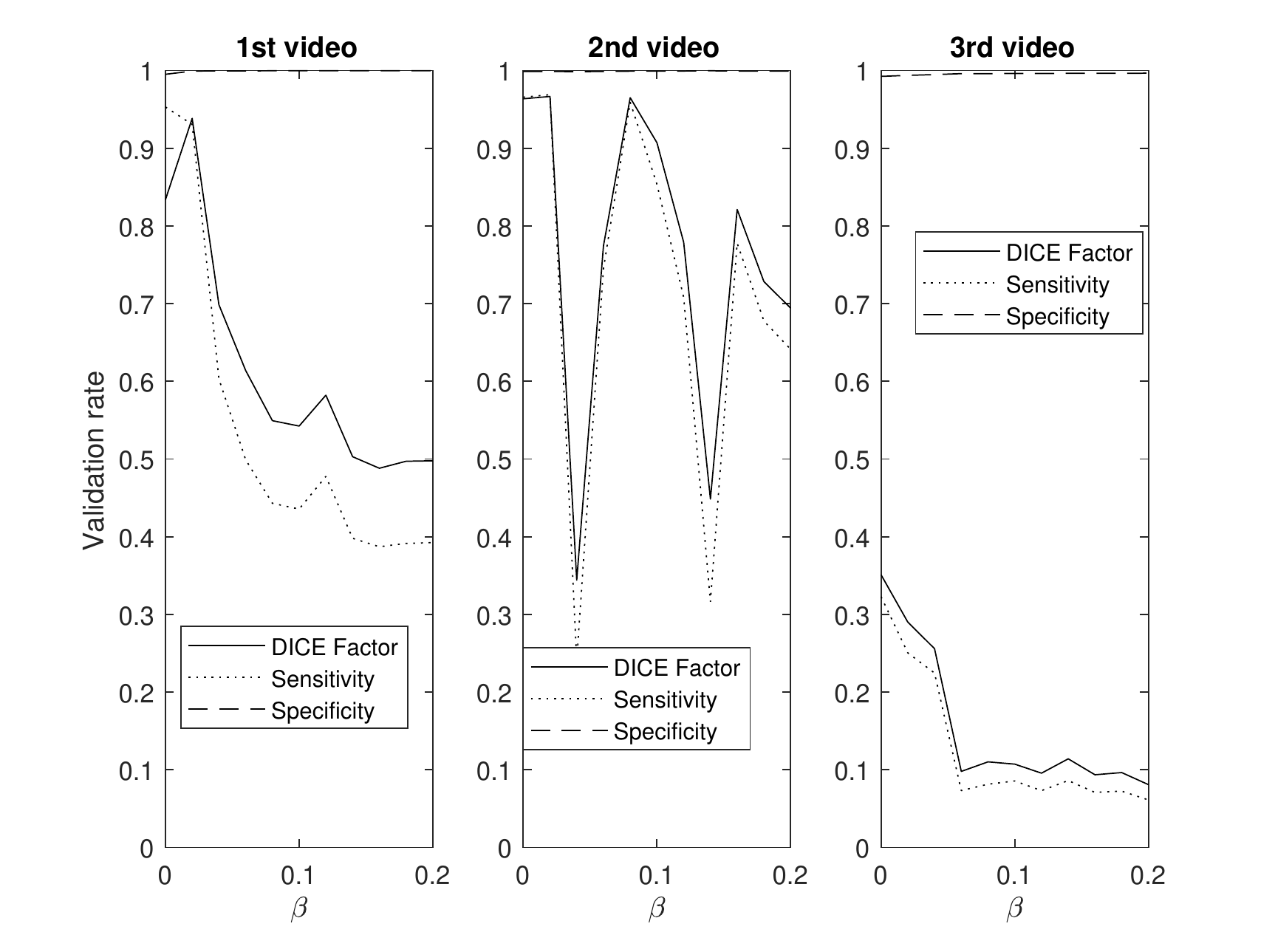} 
\vspace{-0.5cm}
\caption{The validation rates in terms of DF, sensitivity, and specificity versus the parameter $\beta$.}\label{Rbeta}
\end{figure}

\begin{figure}[tb!]
\center
\includegraphics[width=0.9\linewidth]{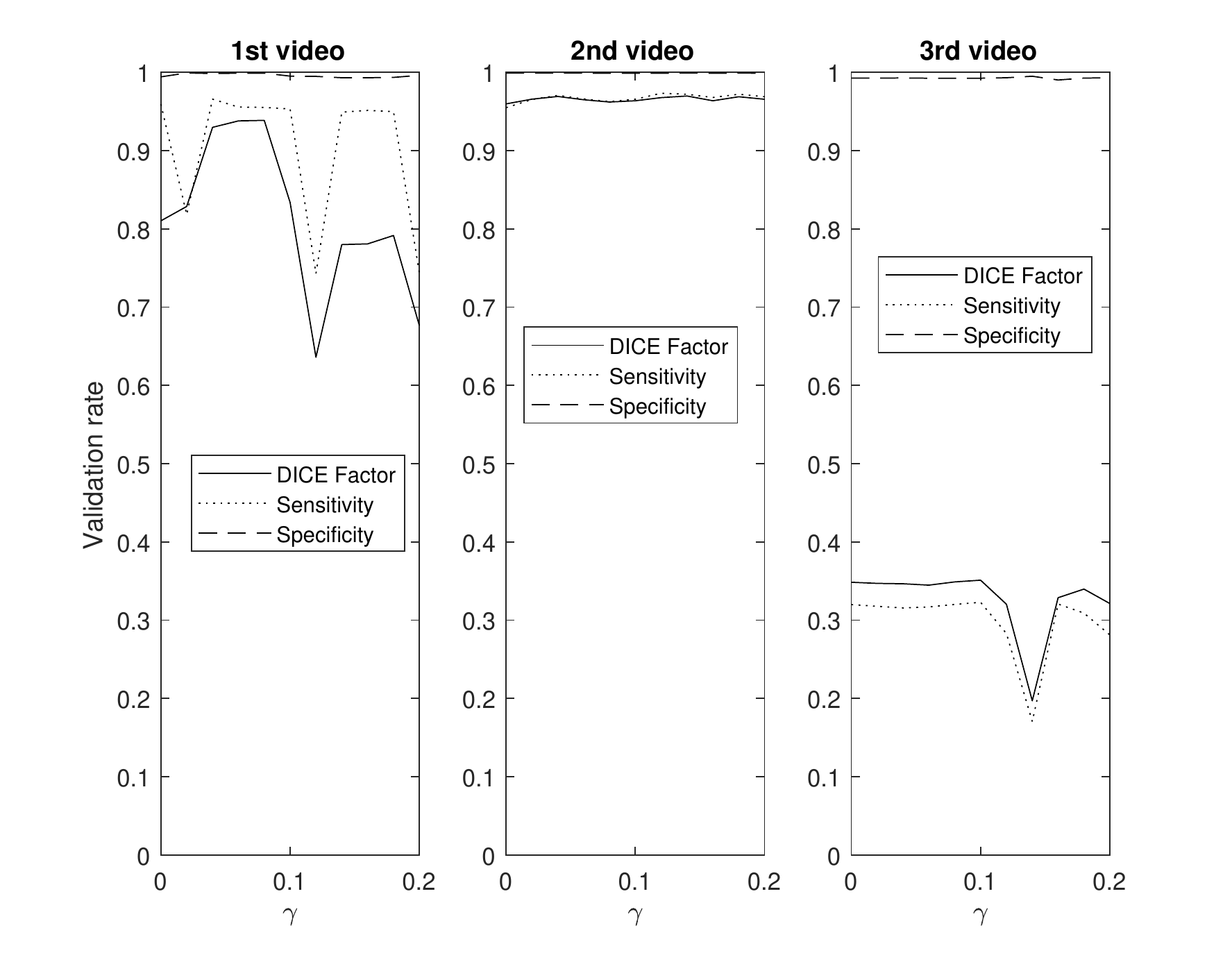} 
\vspace{-0.5cm}
\caption{The validation rates in terms of DF, sensitivity, and specificity versus the parameter $\gamma$.}\label{Rgamma}
\end{figure}

\begin{figure}[tb!]
\center
\includegraphics[width=0.9\linewidth]{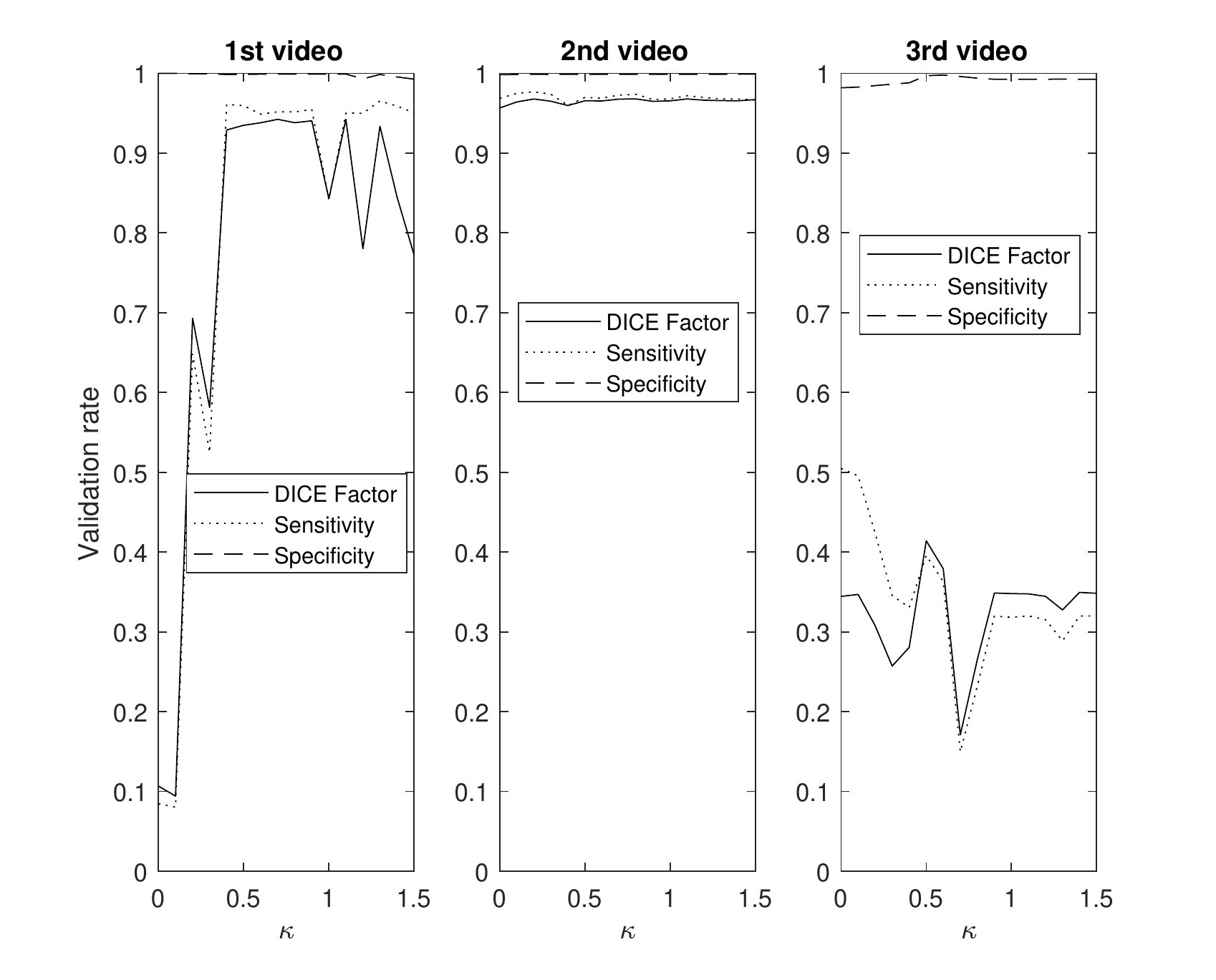} 
\vspace{-0.5cm}
\caption{The validation rates in terms of DF, sensitivity, and specificity versus the parameter $\kappa$.}\label{Rkappa}
\end{figure}

\begin{figure}[tb!]
\center
\includegraphics[width=0.9\linewidth]{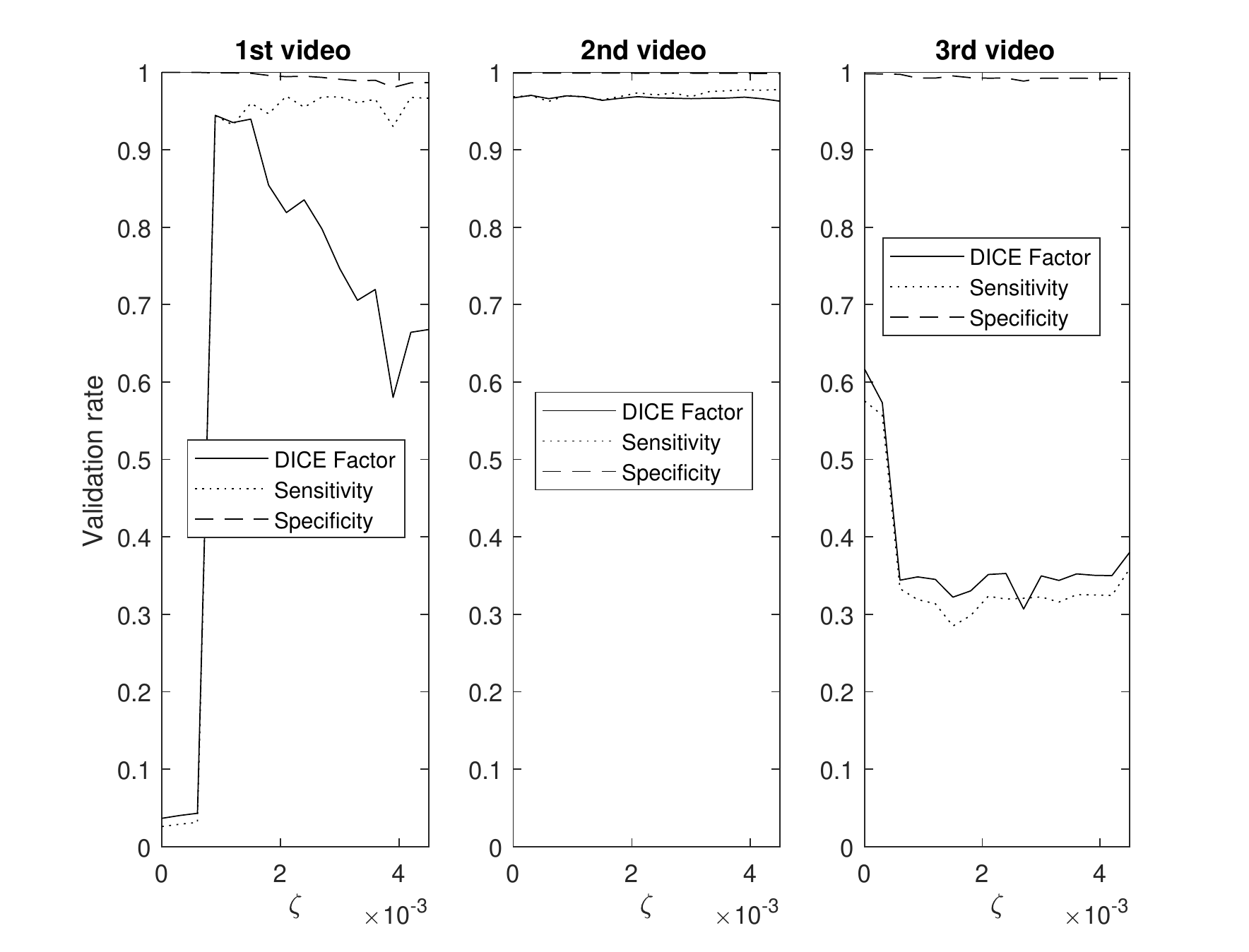} 
\vspace{-0.5cm}
\caption{The validation rates in terms of DF, sensitivity, and specificity versus the parameter $\zeta$.}\label{Rzeta}
\end{figure}

\begin{figure}[tb!]
\center
\includegraphics[width=0.9\linewidth]{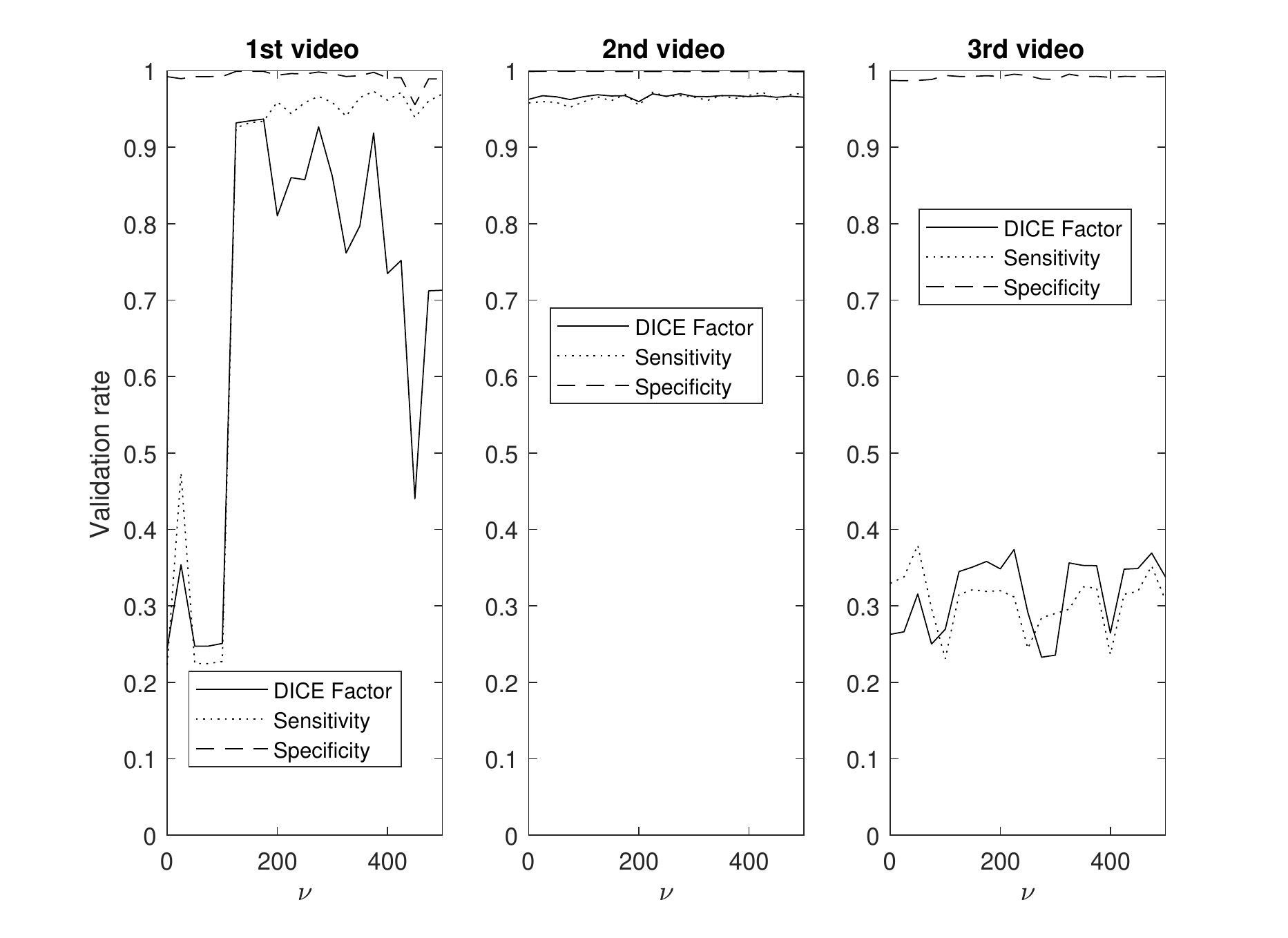} 
\vspace{-0.5cm}
\caption{The validation rates in terms of DF, sensitivity, and specificity versus the parameter $\nu$.}\label{Rnu}
\end{figure}

This section demonstrates the robustness of the algorithm for each parameter along with the relative importance of each parameter on the overall performance. This enables the identification and potential removal of weak features from the energy function in order to improve computational efficiency. For this study, the average DICE factor, sensitivity, and specificity of three different clips versus the initial parameters $\alpha$, $\beta$, $\gamma$, $\kappa$, and $\nu$ are shown in Figs. \ref{Ralpha}-\ref{Rnu}. In the all of these figures, one can easily see that the specificity is always very close to one indicating a relatively small rate of $FP$.
\par
\indent The three test videos suggest setting $\alpha$ to one supports optimal segmentation as shown in Fig. \ref{Ralpha}. Large values of $\alpha$ result in excessive contour shrinking while small values reduce contour smoothness.\\
\indent The parameter $\beta$ demonstrates optimal performance near zero as per Fig. \ref{Rbeta}. This strongly suggests that the continuity energy term is a weak feature and hence, can be removed from the energy function.\\
In Fig. \ref{Rgamma}, all three videos provide their best performance at a $\gamma$ between 0.04 and 0.08, hence, $\gamma$ is set at 0.06. In two of the three test videos, the edge energy does not significantly improve the segmentation performance as the curves appear flat around $\gamma=0$ likely resulting from indistinct edges and consequently, providing limited information to improve segmentation results.\\
\indent Fig. \ref{Rkappa} highlights that different videos demonstrate considerable variance in sensitivities versus $\kappa$. Sensitivity was relatively stable for $\kappa$ ranging between 0.4 and 0.9, hence $\kappa$ was set to 0.8.\\
\indent For the parameter $\zeta$, the best performance was established between 125 - 175 as shown in Fig. \ref{Rzeta}. Hence, $\zeta$ equal to 150 seems to be an appropriate selection.\\
\indent Finally Fig. \ref{Rnu} shows that the best performance is obtained when $\nu$ is between 0.0009 and 0.0015 resulting in $\nu=.0012$ as an appropriate selection.

\subsection{Influence of Contour Points Spacing}
\begin{figure}
\center
\includegraphics[width=0.9\linewidth]{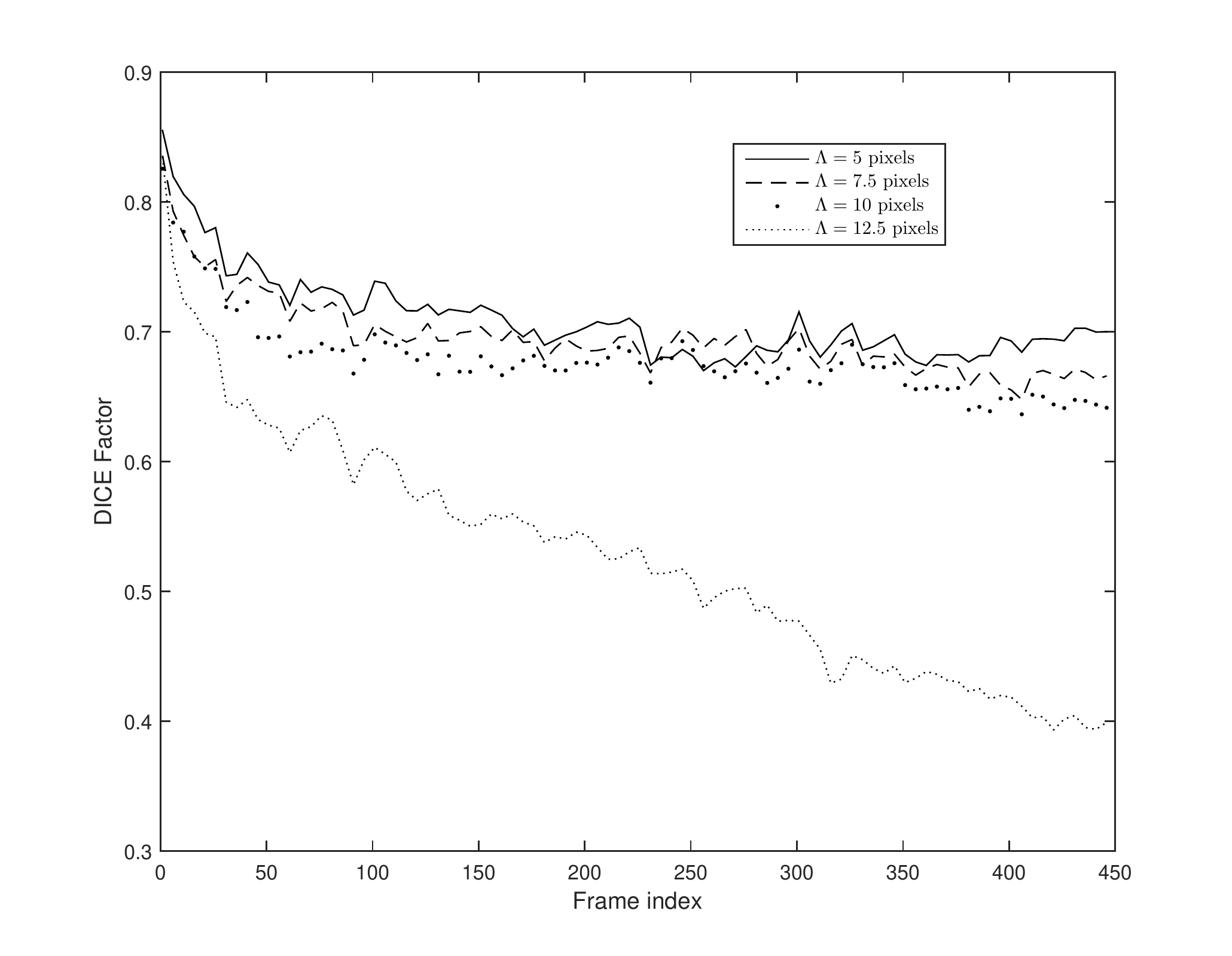} 
\caption{Influence of contour spacing $\Lambda$ on the accuracy of Ad-PAC segmentation.}
\label{figNumber}
\vspace{-0.5cm}
\end{figure}

Here, we study the effect of contour points spacing on the accuracy of the Ad-PAC algorithm. In this study, after segmentation of each frame, the contour is re-sampled and the new value of $N$ is chosen to be $P/\Lambda$, where $P$ is the perimeter of the segmented contour, and $\Lambda$ is the contour points spacing. Fig. \ref{figNumber} presents the average DICE factor obtained from all videos for different values of $\Lambda$. From this figure, one can see that the average DICE factor degrades quickly when the contour spacing is large and it improves when the contour spacing decreases but it nears saturation at $\Lambda=5$ pixels.
\subsection{Tracking Performance}
\begin{figure}[t!]
\centering
\includegraphics[width=0.8\linewidth]{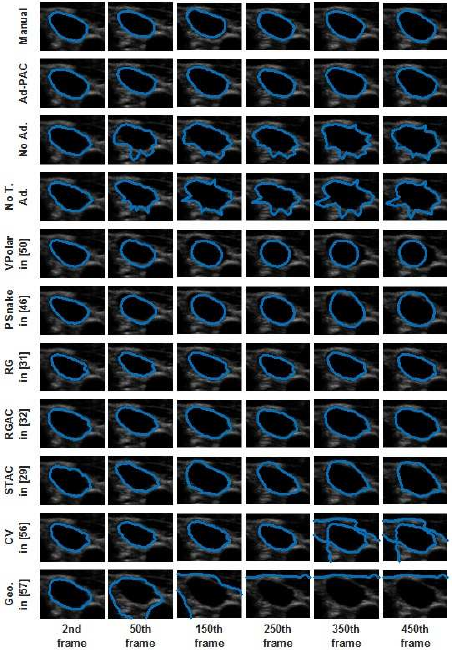}
\caption{Tracking of the IJV in a good quality video for manual segmentation, Ad-PAC and eight other algorithms.}\label{figT1}
\end{figure}

\begin{figure}[t!]
\centering
\includegraphics[width=0.7\linewidth]{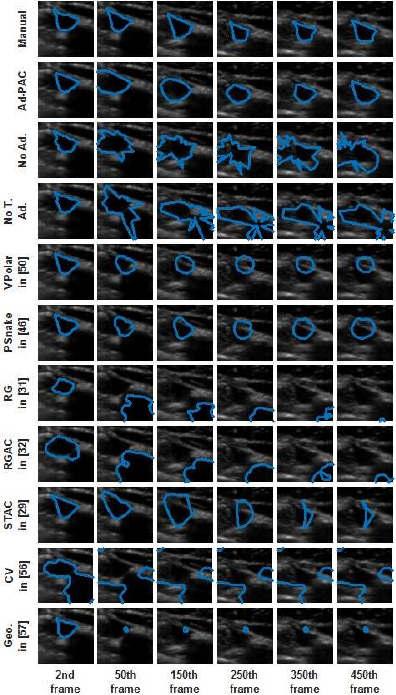}
\caption{Tracking of the IJV in a poor quality video for manual segmentation, Ad-PAC and eight other algorithms.}\label{figT2}
\end{figure}

\begin{figure}[b]
\center
\includegraphics[width=0.9\linewidth]{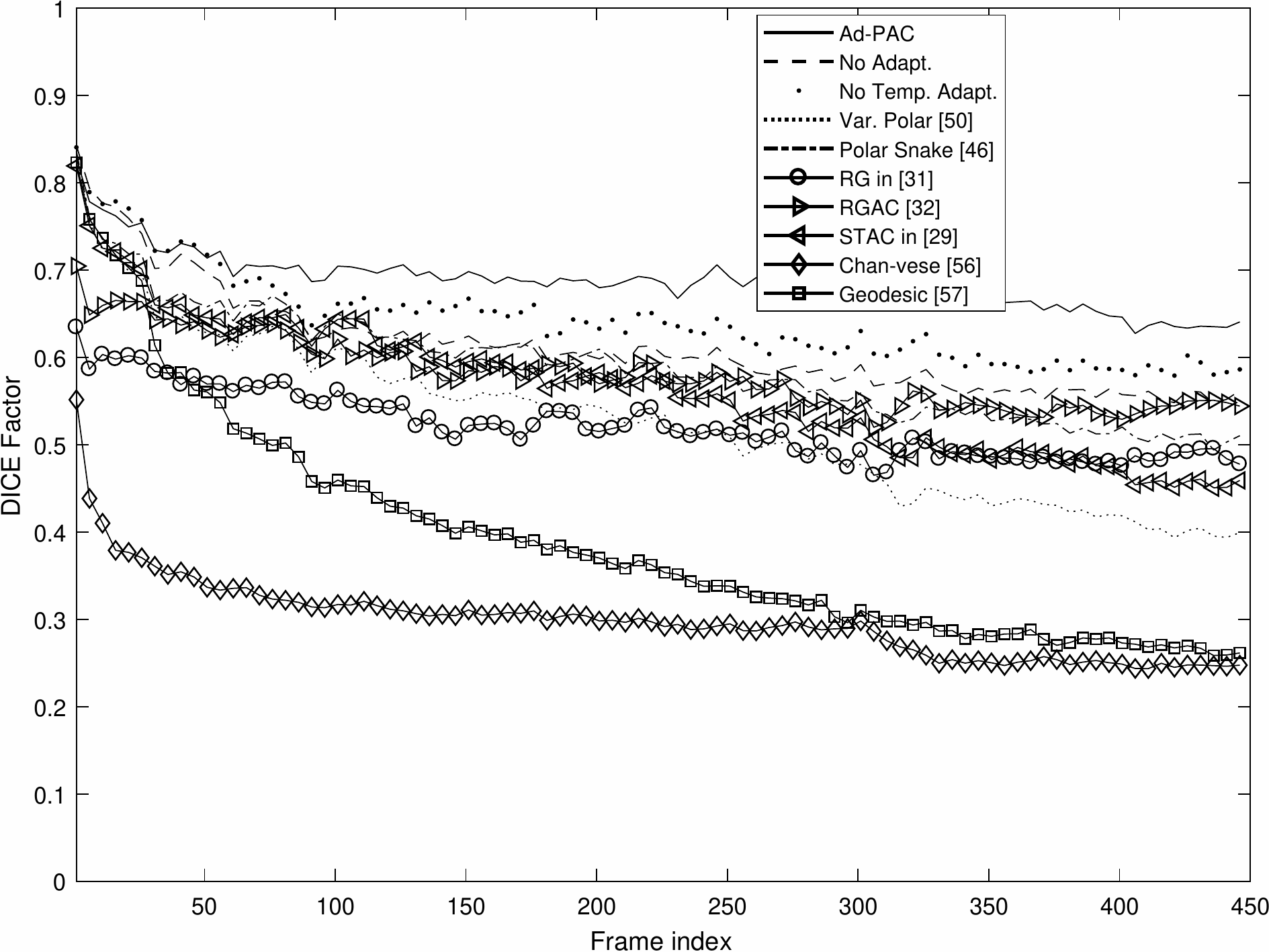} 
\caption{The mean level of agreement between algorithm and manual segmentation versus frame index.}\label{fig_ave}
\end{figure}

\vspace{-0.1cm}
This Section compares the tracking performance of the proposed Ad-PAC algorithm with the manual segmentation and other algorithms as per section V for two sample video as shown in Figs. \ref{figT1} and \ref{figT2}, respectively. From both figures, it is evident that the proposed Ad-PAC algorithm outperforms the existing algorithms and produces results very close to the manual segmentation. Further supporting evidence that parameter adaption significantly improves the performance is evident in rows 3 and 4 row of Figs. \ref{figT1} and \ref{figT2}. The segmented contour is not smooth without parameter adaptation (which is observed as spikes) suggesting that the weight given to the curvature energy term was not sufficiently large enough to compete with the other energy terms and consequently, dominated by them.
Fig. \ref{fig_ave} presents the DICE factors obtained from each algorithm, averaged across all 65 videos irrespective of IJV shape, intensity, speed of variation and quality. From this figure, it is clear that the proposed Ad-PAC algorithm outperforms all existing algorithms with its corresponding DICE factor greater than 0.64. Other algorithms perform significantly worse. In the following sub-sections, more detailed results are presented.
\subsection{Influence of Image Quality}
\begin{figure*}[t!]
\center
\includegraphics[width=0.7\linewidth]{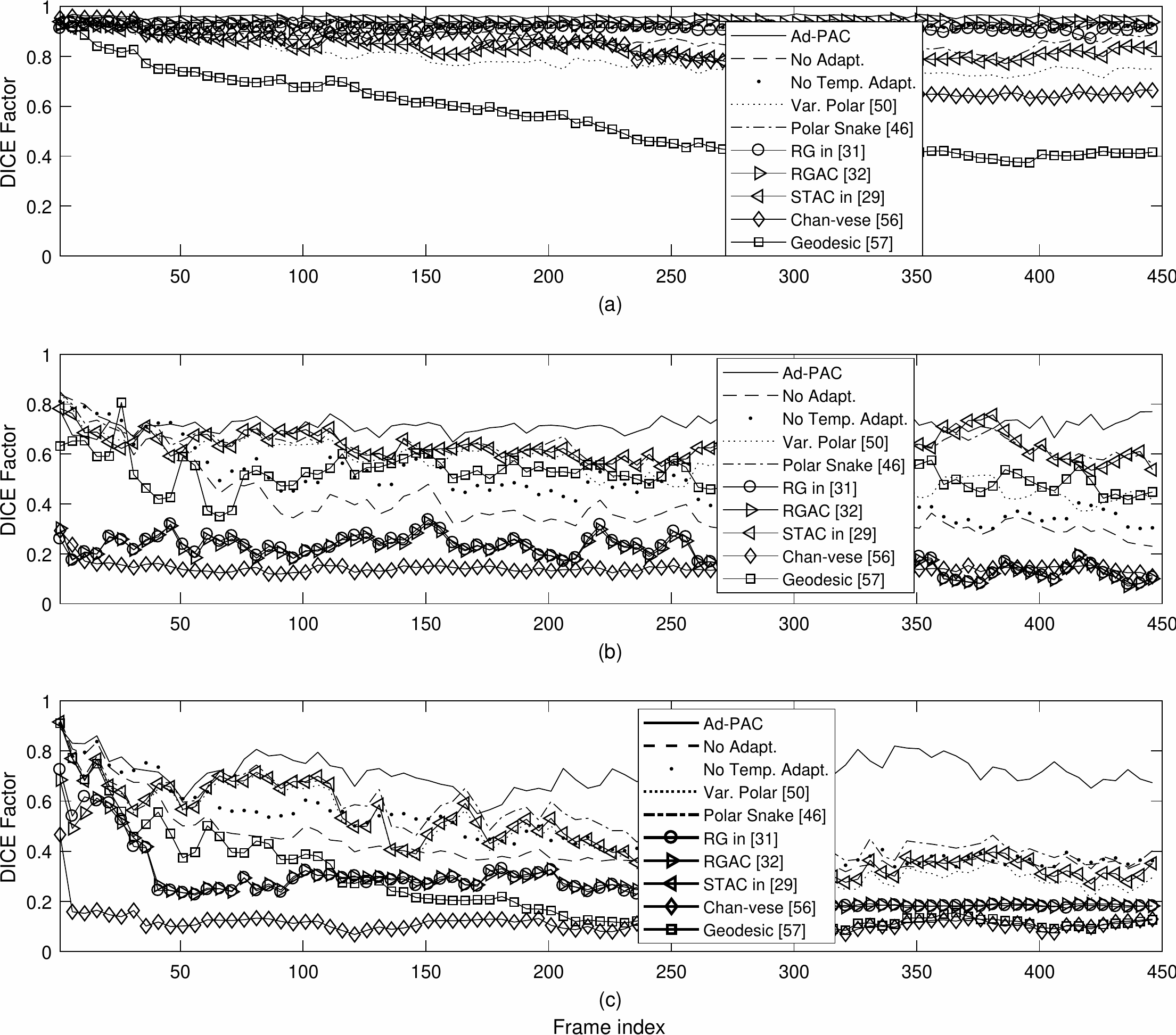} 
\caption{The average level of agreement with manual segmentation versus frame index for ultrasound videos with (a) good, (b) average, and (c) poor qualities.}\label{fig_Q}
\includegraphics[width=0.7\linewidth]{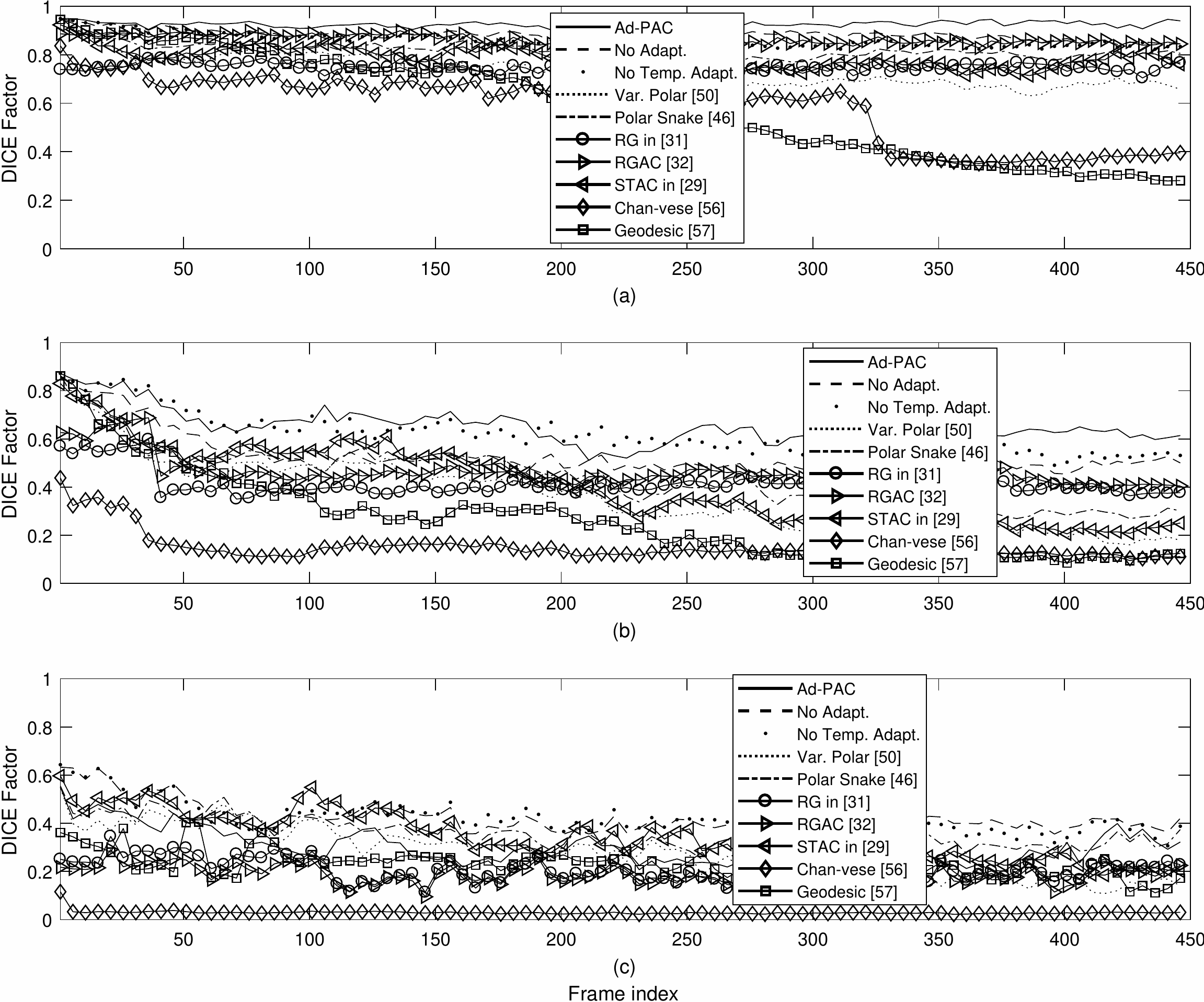} 
\caption{The average level of agreement with manual segmentation versus frame index for (a) oval shape, (b) 1+ apices shape, (c) fully collapsed videos.}\label{fig_S}
\end{figure*}

For this study, all videos were categorized, as good, average, and poor quality videos based on the blinded expert opinion. Fig. \ref{fig_Q} illustrates the DICE results. In good quality ultrasound videos, as per Fig. \ref{fig_Q}-(a), the proposed Ad-PAC algorithm performs very close to the manual segmentation with a DICE factor consistently above 0.95. The minimum value of DICE factors for the other algorithms range from 0.91 down to 0.37 for the Geodesic algorithm \cite{caselles1997}.\\
\indent In average quality videos, as shown in Fig. \ref{fig_Q}-(b), the performance of Ad-PAC algorithm drops as low as 0.65, however, it still outperforms the other AC algorithms. Poor quality videos (Fig. \ref{fig_Q}-(c)) demonstrate the minimum  DICE factor as being 0.55, still above other algorithms.\\

\subsection{Influence of IJV Shape}
The IJV shape also impacts the segmentation results. Oval objects tend to be more suitable for polar representation whereas collapsed vessels tend to be much more challenging. The IJV represents a deformable model influenced by a number of factors including local anatomy, blood volume and blood flow. For this study, IJV videos are categorized into three categories being oval, 1+ apices, and fully collapsed. Fig. \ref{fig_S} presents the average DICE factor for the videos from each category. In Fig. \ref{fig_S}-(a), it is evident that Ad-PAC performs close to manual segmentation when the IJV has an oval shape with a DICE coefficient greater than 0.90. The second best performance belongs to the proposed energy function without temporal adaptation having an average DICE coefficient larger than 0.83. Fig. \ref{fig_S}-(b) shows that the IJV with 1+ apices result in the Ad-PAC performance as low as 0.50 but above other algorithms. Again, the proposed energy function without temporal adaptation is the second best algorithm with a minimum DICE factor as low as 0.48. Is is only when the IJV is fully collapsed does the Ad-PAC algorithm under-perform the algorithm without adaptation and in the worst case, the DICE factor drops to 0.19. This is a limitation of the polar contour model for fully collapsed objects such as the empty IJV.
\subsection{Influence of IJV Variation}
\begin{figure*}
\center
\includegraphics[width=0.7\linewidth]{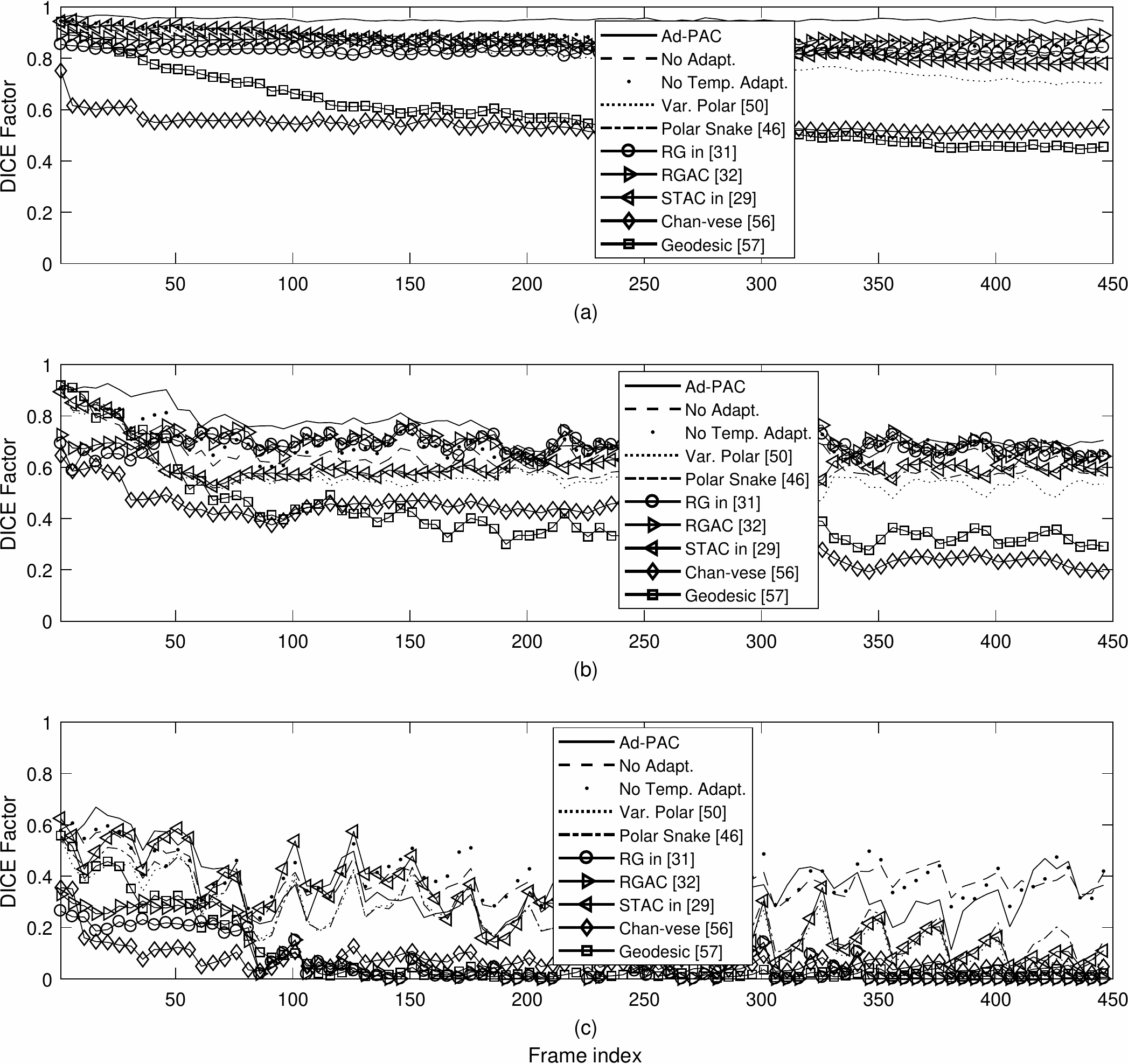} 
\caption{The average level of agreement with manual segmentation for IJV videos with (a) less than 10\% variation, (b) 10-90\% variation, (c) greater than 90\% variation.}\label{fig_Var}
\end{figure*}

The CSA of the IJV undergoes a wide range of variation that typically present a challenge for AC models as they are relatively sensitive to this parameter. This often results in a failure to track and converge to the edges of the object. To study the influence of variation, the ultrasound videos were categorized into three groups i) less than 10 percent, ii) between 10 and 90 percent and iii) more than 90 percent variations. Note that, in the case of more than 90 percent variation, the IJV shape deforms from oval or 1+ apical shape to fully collapsed, resulting in this category resembling the one in Fig. \ref{fig_S}-(c). The numerical results based on this categorization are shown in Fig. \ref{fig_Var}. As one can see from Fig. \ref{fig_S}-(a), when the CSA of the IJV undergoes small variations, the average DICE factor is always greater than 0.94. In Fig. \ref{fig_S}-(b) with the variation between 10 to 90 percent, the Ad-PAC algorithm still performs well with an average DICE factor of 0.64 and still outperforms the other algorithms. Finally from Fig. \ref{fig_S}-(c), it is observed that when the IJV undergoes large variations, all algorithms gradually lose tracking. Ad-PAC algorithm does not always outperform Ad-PAC without adaptation in these scenarios.
\section{Conclusion and Future Work}
In this paper, a novel adaptive polar active contour model (Ad-PAC) is developed for the segmentation and tracking of the internal jugular vein (IJV) in ultrasound imagery. In the proposed algorithm, the parameters of energy function are initialized and locally adapted to the contour features extracted in previous frames. We demonstrate that the extra processing required for parameter adaptation is negligible and that the proposed Ad-PAC algorithm performs well compared with manual segmentation while outperforming multiple existing algorithms across a broad range of image features including image quality, intensity, and temporal variation. 
\par
\indent Although the proposed Ad-PAC algorithm still outperforms existing AC algorithms, the authors intend to address the cases of poor image quality or fully-collapsed IJV by incorporating additional information into its energy function. Furthermore, the focus of this paper centered on the energy function and parameter adaptation with future work being directed at improving the speed and accuracy of the functional minimization through developing more efficient techniques.
\indent Currently, the research team is evaluating the ability of the Ad-PAC algorithm to detect relative changes in circulating blood volume with the intent of predicting when patients with congestive heart failure are at risk of clinical deterioration and subsequent hospitalization.
\clearpage % force a pagebreak
\bibliographystyle{IEEEtran}
%\bibliography{APC}

\begin{IEEEbiography}
[{\includegraphics[width=1in,height=1.25in,clip,keepaspectratio]{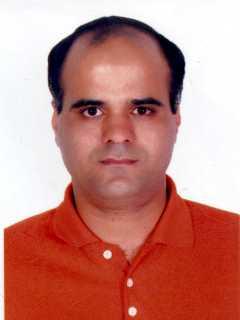}}]{Ebrahim Karami} received the B.S. degree in electrical engineering (electronics) from Iran University of Science and Technology, Tehran, Iran, in 1996 and the M.S. degree in electrical engineering (bioelectric) and the Ph.D. degree in electrical engineering (communications) from the University of Tehran, in 1999 and 2005, respectively. 
Between 2006 and 2012, he was with the Center for Wireless Communications, University of Oulu, Oulu, Finland; Carleton University, Ottawa, ON, Canada; and the University of Saskatchewan, Saskatoon, SK, Canada. He is currently with Memorial University, St John’s, NL, Canada. His current research interests include medical imaging, computer vision, and sensor networks.
Dr. Karami is an associate editor of IET Electronics Letters.
\end{IEEEbiography}
\begin{IEEEbiography}
[{\includegraphics[width=1.25in,height=1in,clip,keepaspectratio]{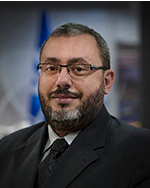}}]{Mohamed Shehata} 
obtained his B.Sc. degree with honors in 1996, his M.Sc. degree in computer engineering in 2001 from Zagazig University, Egypt, and then his Ph.D. in 2005 from the University of Calgary, Canada. Following his Ph.D., he worked as a Post-doctoral Fellow at the University of Calgary on a joint project between the University of Calgary and the Canadian Government, called Video Automatic Incident Detection. After that, he joined Intelliview Technologies Inc. as a Vice-President of the research and development department. In 2013, after seven years in the industry, Dr. Shehata joined Faculty of Engineering and Applied Science at MUN as an Assistant Professor of computer engineering. His research activities include computer vision, image processing, and software design.
\end{IEEEbiography}
\begin{IEEEbiography}
[{\includegraphics[width=1in,height=1.25in,clip,keepaspectratio]{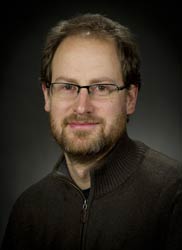}}]{Andrew Smith} is an assistant professor with the Primary Healthcare Research Unit within the Discipline of Family Medicine, Memorial University. He is currently working as a family medicine physician after spending 8 years in the ER. Andrew holds a Master’s degree in Electrical Engineering and holds a cross-appointed to the Faculty of Engineering and Applied Sciences. He is working to create formal Biomedical Engineering opportunities at Memorial. His research interests include Point of Care Ultrasound, Remote Patient Monitoring and Point of Care Genetic Testing technologies. Andrew has recently founded two medical technology start-up companies and is keenly interested in supporting innovation and entrepreneurship within healthcare.
\end{IEEEbiography}
\end{document}